\documentclass[aps, prd, onecolumn, tightenlines, notitlepage, superscriptaddress, nofootinbib, preprintnumbers, floatfix,showkeys,11pt,altaffilletter]{revtex4-2}

\usepackage{upgreek}

\usepackage[official]{eurosym}
\usepackage[normalem]{ulem}
\usepackage{amstext}
\usepackage{graphicx}
\graphicspath{{plots/}}
\usepackage{url}
\usepackage{color}
\usepackage{ulem}
\usepackage[version=4]{mhchem}
\usepackage[utf8]{inputenc}
\usepackage{fontawesome}
\usepackage{yfonts}
\usepackage{amsmath}

\pdfoutput=1
\usepackage{textcomp}
\usepackage{comment}
\usepackage{yfonts}
\usepackage{epsfig,amsfonts,mathrsfs,graphicx,color,slashed,multirow}
\usepackage{latexsym,graphicx,slashed,color,enumerate,url,cancel,gensymb}
\usepackage{textcomp}

\usepackage{slashed}
\usepackage{xspace}

\usepackage[x11names]{xcolor}
\usepackage[colorlinks,pdfstartview=FitV,breaklinks=true]{hyperref}
\usepackage{booktabs}
\usepackage{adjustbox}
\usepackage{textgreek} 
\usepackage{caption, subcaption} 
\usepackage{ragged2e} 
\DeclareCaptionJustification{justified}{\justifying}

\usepackage{booktabs} 
\usepackage{float}
\usepackage{orcidlink}
\usepackage{lmodern}
\usepackage{ae,aecompl}
\usepackage{appendix}
\usepackage{accents}

\usepackage{tikz}
\usetikzlibrary{calc,shapes.geometric,backgrounds,decorations,positioning}

\pgfmathparse{atan2(0,1)}
\ifdim\pgfmathresult pt=0pt 
\tikzset{declare function={atanXY(\x,\y)=atan2(\y,\x);atanYX(\y,\x)=atan2(\y,\x);}}
\else                       
\tikzset{declare function={atanXY(\x,\y)=atan2(\x,\y);atanYX(\y,\x)=atan2(\x,\y);}}
\fi


\def\d{\mathrm{d}}

\definecolor{vdrgreen}{rgb}{0.0, 0.6, 0.0}
\definecolor{persiangreen}{rgb}{0.0, 0.65, 0.58}
\definecolor{mediumpersianblue}{rgb}{0.0, 0.4, 0.65}
\definecolor{dbg}{RGB}{59, 51, 85} 
\definecolor{seaweed}{RGB}{70,129,137}
\definecolor{mintcream}{RGB}{250,247,233} 
\definecolor{lev}{RGB}{193,189,231}
\definecolor{lb}{RGB}{199,223,226}
\definecolor{qp}{RGB}{235,198,203}
\definecolor{dm}{RGB}{142, 50, 61}
\definecolor{pumpkin}{RGB}{254, 127, 45}
\definecolor{av}{RGB}{169, 140, 190}
\definecolor{iris}{RGB}{88, 80, 190}
\definecolor{ultraviolet}{RGB}{95, 90, 162}
\definecolor{burntsienna}{RGB}{231, 111, 81}
\definecolor{celestialblue}{RGB}{16, 151, 214}
\definecolor{crayola}{RGB}{255, 89, 100}

\makeatletter
    \newcommand{\colorboxed}[3][white]{\fcolorbox{#2}{#1}{\m@th$\displaystyle#3$}}
\makeatother

\AtBeginDocument{\hypersetup{citecolor=mediumpersianblue,linkcolor=mediumpersianblue,urlcolor=mediumpersianblue}}
\usepackage{appendix}

\newcommand{\BIG}{{\sf BIG}}
\newcommand{\SLIM}{{\sf SLIM}}
\newcommand{\QUAINT}{{\sf QUAINT}}
\newcommand{\usine}{{\sc usine}}

\begin{document}


\preprint{}

\title{{\Large 
Antinuclei from Primordial Black Holes}}

\author{Valentina De Romeri~\orcidlink{0000-0003-3585-7437}}
\email{deromeri@ific.uv.es}

\affiliation{Instituto de F\'{i}sica Corpuscular (CSIC-Universitat de Val\`{e}ncia), Parc Cient\'ific UV C/ Catedr\'atico Jos\'e Beltr\'an, 2 46980 Paterna (Valencia) - Spain}

\author{Fiorenza Donato~\orcidlink{0000-0002-3754-3960}}
\email{fiorenza.donato@unito.it
}

\affiliation{Dipartimento di Fisica, Università di Torino, and INFN, Sezione di Torino
Via P. Giuria 1, Torino, Italy }

\affiliation{Theoretical Physics Department, CERN,  Esplanade des Particules 1, CH-1211 Geneva 23, Switzerland}
\author{David~Maurin~\orcidlink{0000-0002-5331-0606}}
\email{dmaurin@lpsc.in2p3.fr
}
\affiliation{LPSC, Université Grenoble-Alpes, CNRS/IN2P3, 38026, Grenoble, France}

\author{Lorenzo Stefanuto~\orcidlink{0009-0003-9377-8654}}
\email{lorenzo.stefanuto@unito.it
}
\affiliation{Dipartimento di Fisica, Università di Torino, and INFN, Sezione di Torino
Via P. Giuria 1, Torino, Italy }

\author{Agnese Tolino~\orcidlink{0009-0003-3278-0902}}
\email{atolino@ific.uv.es}
\affiliation{Instituto de F\'{i}sica Corpuscular (CSIC-Universitat de Val\`{e}ncia), Parc Cient\'ific UV C/ Catedr\'atico Jos\'e Beltr\'an, 2 46980 Paterna (Valencia) - Spain}

\keywords{primordial black holes, Hawking evaporation, cosmic rays, antimatter}

\begin{abstract}
Light primordial black holes (PBHs) may have originated in the early Universe, and could contribute to the dark matter in the Universe. 
Their Hawking evaporation into particles could eventually lead to the production of antinuclei, which 
propagate and arrive at Earth as cosmic rays with a flux peaked at GeV energies. 
We revisit here the antiproton and antideuteron signatures from PBH evaporation, relying on a lognormal PBH mass distribution, state-of-the-art propagation models, and an improved coalescence model for fusion into antideuterons. 
Our predictions are then compared with AMS-02 data on the antiproton flux. 
We find that the AMS-02 antiproton data severely constrain the Galactic PBH density, setting bounds that depend significantly on the parameters of the 
lognormal mass distribution, and that are comparable to or slightly stronger than bounds set from diverse messengers. 
 We also discuss prospects for future detection of antideuterons. Given the bounds from AMS-02 antiproton data, we predict that if antideuterons were to be measured by AMS-02 or GAPS, since the secondary contribution is subdominant, they would clearly be a signal of new physics, only part of which could, however, be explained by PBH evaporation. 

\end{abstract}
\maketitle

\section{Introduction}
Primordial black holes (PBHs) are a hypothetical form of black holes~\cite{Hawking:1971ei,Zeldovich:1967lct,Carr:1974nx,Chapline:1975ojl} that may have originated in the early Universe, via a wide range of different mechanisms. One motivated formation scenario is from the collapse of large primordial inhomogeneities, before the Big-Bang nucleosynthesis epoch. 
PBHs have garnered renewed interest in recent years, after the observation of gravitational waves in the LIGO-Virgo-KAGRA experiments~\cite{Abbott:2016blz,Abbott:2020gyp,KAGRA:2021vkt}. These observations, while confirming the existence of astrophysical black holes, also provide new opportunities to investigate a possible primordial origin of the observed events.

PBHs also constitute one of the earliest proposed candidates to explain the unknown dark matter (DM) component of the Universe~\cite{Aghanim:2018eyx,Bertone:2016nfn,Cirelli:2024ssz}. We refer the reader to  Refs.~\cite{Carr:2020xqk,Bird:2022wvk,Green:2024bam} for recent reviews on PBHs as DM. Currently, only PBHs with masses in the range $10^{17} \rm g \lesssim M_{\rm PBH} \lesssim 10^{22}$ g could explain the totality of DM~\cite{Carr:2016hva,Green:2020jor,Carr:2020xqk}, a range usually referred to as the \textit{asteroid-mass} region. Lighter PBHs are expected to evaporate, emitting fluxes of fundamental particles with a quasi-thermal spectrum, as originally predicted by Hawking~\cite{Hawking:1974rv,Hawking:1974sw}. PBHs lighter than $M_{\rm PBH} \lesssim 5 \times 10^{14}$ g would have a lifetime shorter than the age of the Universe~\cite{Page:1976df,Page:1976ki,MacGibbon:2007yq} and, therefore, could not contribute to the observed DM abundance. \\

The emission of particles through Hawking evaporation provides plenty of opportunities to probe PBHs in their light-mass regime. Under the assumption that the semi-classical approximation is valid up to close to the Planck scale, the evaporation proceeds slowly in its first stage. Then, as the PBH mass decreases, its temperature increases, leading to an accelerated evaporation process. The final moments of the black hole life would resemble an explosion~\cite{Hawking:1974rv}. 
The non-observation of the PBH evaporation products allows to set stringent constraints on the expected abundance of these objects~\cite{Carr:2020gox,Auffinger:2022khh}. Different messengers have been considered in this scope, like photons (mainly
$\gamma$ rays and X rays)~\cite{1976ApJ...206....8C,Carr:2009jm,Lehoucq:2009ge,Wright:1995bi,Arbey:2019mbc,Ballesteros:2019exr,Laha:2020ivk,Tan:2024nbx}, antimatter like electrons/positrons~\cite{Boudaud:2018hqb,Dasgupta:2019cae,DeRocco:2019fjq,Laha:2019ssq,DelaTorreLuque:2024qms}, and neutrinos~\cite{Halzen:1995hu,Lunardini:2019zob,Dasgupta:2019cae,Wang:2020uvi,DeRomeri:2021xgy,Bernal:2022swt,DeRomeri:2024zqs}.

Heavier charged particles, such as protons/antiprotons, and heavier nuclei and antinuclei can also be emitted as products of the evaporation process, following the hadronization of the particles directly emitted through Hawking radiation
\cite{Barrau:2001ev,Barrau:2002mc,Maki:1995pa,Herms:2016vop}.  Cosmic-ray (CR) antinuclei thus constitute another promising avenue to search for signatures of PBH evaporation. The antiproton flux predicted from Hawking evaporation of PBHs is expected to peak at lower energies, compared to the secondary antiprotons produced by spallations of CR nuclei onto the interstellar medium (ISM)~\cite{Barrau:2001ev}. Sub-GeV and GeV antiproton data are therefore particularly suited to probe exotic origins such as PBH evaporation.

The data from CR antinuclei are much more limited than those from nuclei, due to their suppressed production rate. Thanks to recent space-based experiments, CR antiprotons have been measured on a wide energy range by the satellite experiment PAMELA \cite{Adriani:2008zq,PAMELA:2010kea}, 
and an even wider range and higher precision by AMS-02 on the International Space Station \cite{AMS:2016oqu,AMS:2021nhj}. 
CR antideuterons have a suppressed production rate compared to antiprotons (see next), and only upper limits on their flux have been derived so far, the best ones being those of the balloon-borne BESS Polar-II instrument \cite{BESS:2024yma}. 
In the near future, the GAPS experiment \cite{Aramaki_2016,GAPS:2023hag} will undertake a series of long-duration Antarctic balloon flights. The detector is optimized for antideuterons, having a sensitivity about two orders of magnitude better than the current BESS limits \cite{GAPS:2023hag}, but it will also measure antiprotons with unprecedented precision in the low-energy range \cite{GAPS:2023hag,GAPS:2022ncd}.\\

In this work, we revisit the antiprotons and antideuterons signatures from PBH evaporation. PBHs as the primary
source of CR antiprotons were first studied in Ref.~\cite{Kiraly:1981ci} (based on outdated Galactic propagation models) and then further developed in Ref.~\cite{Turner:1981ez}. 
Following studies, based on the improved treatment of secondaries~\cite{MacGibbon:1991tj,MacGibbon:1990zk}  from Hawking radiation~\cite{MacGibbon:1991vc}, exploited the precise low-energy antiproton measurements by the BESS experiment~\cite{Maki:1995pa,Mitsui:1996qy,2012PhRvL.108e1102A,Yamamoto:2013yva} to set more stringent limits on the density of PBHs, or provided sensitivities for GAPS~\cite{Aramaki:2014oda}. To increase the sensitivity of CR experiments to PBHs, Ref.~\cite{Wells:1998jv} even proposed sending an instrument outside the solar cavity, where low-energy CR fluxes are not suppressed by solar modulation. More realistic estimates of the antiproton flux from PBH evaporation were given in Refs.~\cite{Barrau:2001ev,Barrau:2002ru}, relying on updated propagation models tuned on wider and more precise CR data, and assuming an extended mass function for the PBH population. 
Regarding the possible formation of antideuterium nuclei from PBH evaporation, the first calculation of the expected flux was given in Ref.~\cite{Barrau:2002mc} and then refined in Refs.~\cite{Aramaki:2015pii,Herms:2016vop}. The most recent studies on antiprotons and antideuterons, in relation to PBH physics, are discussed
in Refs.~\cite{Herms:2016vop,Serksnyte:2022onw}. A possible connection to anomalies in astrophysical data has recently been investigated in Ref.~\cite{Korwar:2024ofe}.\\

Here, we build upon these previous works and update them in several ways. First, we evaluate the expected fluxes of antiprotons and antideuterons  from PBH evaporation assuming, for the first time, an extended lognormal mass distribution. Next, we use state-of-the-art models for Galactic propagation, relevant for a realistic evaluation of the detectable spectra of CR antinuclei. Finally, we analyze AMS-02 antiproton data to obtain updated constraints on the PBH density in the Galaxy, assuming different widths for the initial PBH mass distribution.  We further translate such bounds into the fraction of DM that can be composed of PBHs and into the local PBH explosion rate, under some assumptions. We finally discuss prospects for the detection of CR antideuterons from PBH evaporation with GAPS and AMS-02.\\

Our paper is organized as follows. We present in Sec.~\ref{sec:evaporation} our computation of the antinuclei source spectra from Hawking radiation of PBHs. We discuss the propagation models used in our analysis in Sec.~\ref{sec:fluxes} and provide the expected antinuclei fluxes at Earth. The statistical analysis of AMS-02 antiproton data is presented in  Sec.~\ref{sec:pbar}. We discuss perspectives with antideuteron measurements in Sec.~\ref{sec:dbar},  and conclude in Sec.~\ref{sec:conclusions}.

\section{Evaporation of PBHs into antimatter}
\label{sec:evaporation}

PBHs may have formed in the early Universe via a wide range of possible mechanisms~\cite{Escriva:2022duf}. One of the first proposed scenarios postulates that PBHs could have formed during the radiation-dominated era, from the collapse of large primordial inhomogeneities~\cite{Carr:1974nx,Carr:1975qj}. 

Given the nature of the gravitational collapse leading to their formation, PBHs are expected to span an extended range of masses~\cite{Niemeyer:1997mt}. 
We can define the following quantity to account for the PBH number density per initial mass ($M_{\rm in}$) interval,

\begin{align} 
 g(r, z, M_{\rm in})\equiv M_{\rm in}\frac{\d n_{\rm PBH}}{\d M_{\rm in}} = M_{\rm in} \frac{\d^2 N_{\rm PBH}}{\d M_{\rm in}  \d V} \, ,
        \label{eq:g(M)}
\end{align}
where $n_{\rm PBH}$ is the PBH number density, $\d V$ is the volume interval and $N_{\rm PBH}$ refers to the overall number of PBHs.
Assuming a realistic production mechanism $ g(r, z, M_{\rm in})$ can be described phenomenologically with a lognormal distribution~\cite{Dolgov:1992pu,Green:2016xgy}, dubbed \textit{ln},

\begin{align} 
\left.  g(r, z, M_{\rm in})\right|_{\rm ln}  = 
        \rho_{\rm PBH}(r,z) \frac{\mathcal{A}}{\sqrt{2 \pi} \sigma M_{\rm in}} \exp{ \left[  - \frac{\log^2(M_{\rm in}/\mu_c)}{2 \sigma^2}\right] } \, ,
        \label{eq:lognormal}
\end{align}
with $\mu_c$ the critical mass at which the function $g(M_{\rm in})$ peaks, $\sigma$ the width of the distribution, and $\rho_{\rm PBH}(r,z)$ the PBH mass density at the cylindrical coordinates $r$ and $z$ (that describe the position in our Galaxy).
The distribution is normalized via the normalization constant $\mathcal{A}$, through the condition $\int \d M g(r,z,M) = \rho_{\rm PBH}(r,z)$. 
The lognormal mass function in Eq.~\eqref{eq:lognormal} has been shown to be a good approximation when PBHs form from a symmetric peak in the inflationary power spectrum~\cite{Kannike:2017bxn,Green:2016xgy}. 
Different production mechanisms can envisage different initial mass distributions. As an example, PBHs formed from scale-invariant primordial density fluctuations or from the collapse of cosmic strings are well described by an initial mass spectrum that follows a power-law distribution~\cite{Carr:1975qj}, dubbed \textit{pow},
\begin{align} 
\left. g(r, z, M_{\rm in}) \right|_{\rm pow} =  \rho_{\rm PBH}(r,z) \mathcal{A}
        M_{\rm in}^{-3/2} \, .
        \label{eq:powerlaw}
\end{align}
Alternatively, a scenario in which all PBHs have one single mass $M_{\rm mc}$, usually dubbed monochromatic (\textit{mc}) mass function, has also been considered as a convenient and simplified approximation. In this case, $g(r,z, M_{\rm in})$ is described by a Dirac delta function
\begin{align} 
\left. g(r, z, M_{\rm in}) \right|_{\rm mc}  =  \rho_{\rm PBH}(r,z) \delta(M_{\rm mc} - 
        M_{\rm in}) \, .
        \label{eq:monochromatic}
\end{align}

In our analysis, we will mainly consider a lognormal distribution and present our results in terms of a few reference values for its width parameter $\sigma$. Indeed, compared to the simple monochromatic case (see, for instance, Refs.~\cite{Carr:2017jsz,Kuhnel:2017pwq}), several works have already shown how considering extended mass distribution dramatically affects the phenomenological bounds that can be inferred.

\subsection{Hawking radiation}

According to Hawking's predictions, black holes are expected to radiate thermally with a temperature\footnote{Throughout this paper, we consider natural units, i.e., $\hbar = c = k_{\rm B} = 1$, and define the Planck mass as $M_{\rm pl}=1/\sqrt{G_N}$, where $G_N$ is the gravitational constant.}~\cite{Hawking:1974rv,Hawking:1974sw}
\begin{equation}
T = \frac{1}{8 \pi G_N M},
\end{equation}
$M$ being the black hole mass and $T$ its temperature.
PBHs are hence predicted to evaporate, with a particle emission rate given by 
\begin{equation}
    \frac{\d M}{\d t} = - \frac{\alpha\left(M\right)}{M^2},
    \label{eq:loss_rate}
\end{equation}
if we assume that the semi-classical picture is valid up to close to the Planck scale.  In the following, we will always assume this approximation to hold throughout the whole PBH lifetime~\cite{Hawking:1974sw}. It has recently been noted, however, that this approach might not be self-consistent, as it neglects the back-reaction of the emission on the quantum state of the black hole. The inclusion of a \textit{memory-burden} effect~\cite{Dvali:2018xpy,Dvali:2020wft,Dvali:2024hsb} is currently being discussed, as it would affect the evaporation process, slowing it down and potentially extending the PBH
lifetime; see, for example, Refs.~\cite{Alexandre:2024nuo,Thoss:2024hsr,Montefalcone:2025akm,Dvali:2025ktz}. 
Going back to Eq.~\eqref{eq:loss_rate}, $\alpha\left(M\right)$ is the evaporation coefficient that depends on the mass of the PBH itself, and encodes all emitted particle species with energy $E$
and their degrees of freedom~\cite{MacGibbon:1990zk,MacGibbon:1991tj}:
\begin{equation}
\label{eq:alpha}
\alpha\left(M\right) = M^2 \sum_i\int_0^\infty\frac{\d^2N_i}{\d E \d t} E \d E\, .
\end{equation}
The function $\alpha\left(M\right)$ is therefore a combination of step functions, increasing with decreasing PBH masses. Indeed, as the PBH radiates, it loses mass and gets hotter; any time the PBH temperature increases above a particle species mass a new degree of freedom becomes accessible in the evaporation process. Counting all fundamental SM particles, the evaporation coefficient eventually reaches the asymptotical value $\alpha_{\rm SM} \simeq 8 \times 10^{26}$\,g$^3$\,s$^{-1}$~\cite{MacGibbon:1991tj,Ukwatta:2015iba,Baker:2021btk} for $M \lesssim 10^{10}$\,g or, equivalently,  $T \gtrsim 1$\,TeV.

The instantaneous differential emission rate by Hawking radiation (HR) per fundamental particle species reads 
\begin{align}
\label{eq:dNdEdt_pri}
    \left.\frac{\d^2N}{\d E \d t}\right|_{\rm HR}\left(E, M \right) = \frac{g}{2\pi} \frac{\Gamma_{s}(E)}{ {\exp(E/T)} - (-1)^{2s}} \, ,
\end{align}
where $g$ are the internal degrees of freedom associated to the particle  with spin $s$, and $E$ refers to the total energy of the emitted particle. 
Equation~\eqref{eq:dNdEdt_pri} describes a Fermi-Dirac or Bose-Einstein thermal spectrum, depending on the emitted particle's spin $s$, corrected by the greybody factors
$\Gamma_{s}(E)$, which quantify the possible back-scattering of particles and, as a consequence, the departure from a black-body spectrum. We will refer to the direct emission of fundamental particles in the evaporation process as to the HR spectrum.

The evaporation process is particularly efficient for light PBHs which, due to the radiation of fundamental particle species, can lose a large amount of mass or, eventually, evaporate completely. As a consequence, we expect the PBH mass distribution to evolve, from their formation time till now, in a non-trivial way, especially for extended distributions like the lognormal one~\cite{Mosbech:2022lfg}.
The PBH number density mass distribution {\em today} is related to the {\em initial} one through

\begin{equation}
    \frac{\d n_{\rm PBH}}{\d M} = \frac{\d n_{\rm PBH}}{\d M_{\rm in}}\frac{\d M_{\rm in}}{\d M} = \frac{\d n_{\rm PBH}}{\d M_{\rm in}} \frac{M^2 \alpha(M_{\rm in})}{M_{\rm in}^2 \alpha(M)}\, ,
\label{eq:distribution_evolution}
\end{equation}
where the relation between $M_{\rm in}$ and $M$ can be found by numerically solving Eq.~\eqref{eq:loss_rate}.

\begin{figure}[t]
    \centering
        \includegraphics[width=0.9\textwidth]{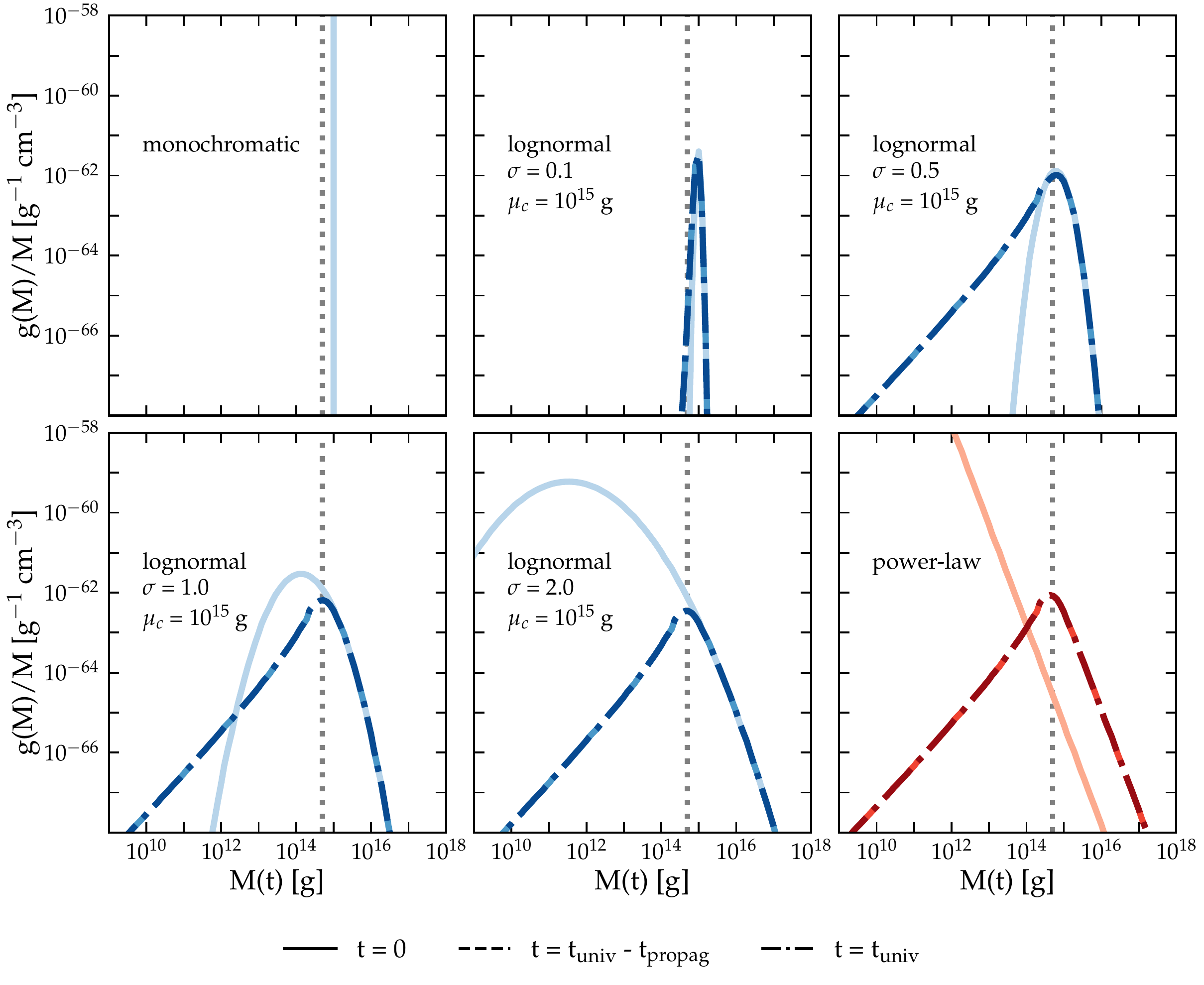}
\caption{Monochromatic (top-left panel) and extended PBH mass distributions (all other panels), $g(M)/M$, evolved at different times. The lognormal distribution, Eq.~\eqref{eq:lognormal}, is shown in blue at fixed $\mu_c$ = $10^{15}$\,g for different widths $\sigma$. The last panel, in red, shows the evolution of a power-law distribution, Eq.~\eqref{eq:powerlaw}. In all panels, the solid light curve denotes the distribution at the time of formation. The dot-dashed (dashed) line identifies the same distribution evolved until the age of the Universe (subtracted by the average propagation time of CR antiprotons). The vertical dotted gray line marks the initial mass $M_{\rm in}^*= 5 \times 10^{14}$ g for a PBH that is evaporating completely today.}
\label{fig:dndM}
\end{figure}
We show in Fig.~\ref{fig:dndM} the PBH number density distribution, $g(M)/M$, for various PBH mass distributions at different times. In all panels, the dotted gray vertical line indicates the critical mass $M_{\rm in}^* = 5 \times 10^{14}$\,g at formation, corresponding to a PBH evaporating completely today. 
The top-left panel represents the monochromatic case, see Eq.~\eqref{eq:monochromatic}, with $M_{ \rm mc} = 10^{15}$\,g. The subsequent four panels show the lognormal distribution case with $\mu_c = 10^{15}$\,g, see Eq.~\eqref{eq:lognormal}, for four different values of the distribution width ($\sigma = 0.1, 0.5, 1.0$, and $2.0$). In all these panels, the solid light blue curves indicate the mass distribution at the time of formation ($t \simeq 0$\,s, in which case $M = M_{\rm in}$ on the $x$-axis), while the blue dot-dashed lines represent the same mass distribution evolved until today ($t = t_U\simeq 4 \times 10^{17}$\,s, in which case $M \equiv M [t_U] $ on the $x$-axis); we also show the distributions as they were a few hundreds of Myr ago (dashed blue lines), to account, with a large safety margin, for the time passed between the production and transport of $\bar{p}$ for the observer, but no visible difference can be seen\footnote{ Notice that since the PBH distribution does not significantly change during the average $\overline{p}$ propagation time, we can consider the source term to be stationary during this time frame.}.
The bottom-right panel in Fig.~\ref{fig:dndM} shows the evolution of a power-law mass distribution, see Eq.~\eqref{eq:powerlaw}: the solid orange line represents the power-law mass distribution at formation time, while the red dot-dashed curve is evolved at the age of the Universe. 
All these mass spectra increase as $M^2$ for PBH masses  smaller than $M_{\rm in}^*$, as previously noted in Refs.~\cite{Barrau:2001ev,Herms:2016vop}.

From these figures, it is evident that, once evolved, the extended distributions 
look similar at present time, especially the slope of their low-mass tails, though their magnitudes differ. Notably, for $\mu_c = 10^{15}$\,g, only widths with $\sigma \gtrsim 0.5$ yield distributions of comparable magnitude.
 As already noted in Refs.~\cite{Barrau:2001ev} and~\cite{Herms:2016vop}, the time evolution causes the lognormal and power-law distributions to develop a tail below $M \lesssim 10^{15}$\,g while, above this mass, no significant changes are induced by the evaporation process.
 Given the rest masses of antinuclei, we expect the greatest contribution to the antinuclei production from PBH evaporation to arise from the low-mass tails of the evolved distributions, $M \lesssim 10^{13}$\,g.

\subsection{Antimatter source spectra from Hawking radiation}
\label{sec:spectra}

Black holes do not directly emit antinuclei ~\cite{MacGibbon:1990zk,Barrau:2001ev,Herms:2016vop,Auffinger:2022ive}. Indeed, quarks, gluons, and other elementary particles produced from PBH evaporation can subsequently hadronize into composite states. When the PBH temperature goes above the quantum chromodynamics confinement scale, $\Lambda_{\rm QCD}$, only fundamental particles are radiated. Note that, for PBHs with $T \lesssim \Lambda_{\rm QCD}$, the emission of fundamental particles is expected to cease, giving way to the direct emission of pions, translating into a sharp drop of up to a factor 5 in $\alpha(M)$ at $M_{\rm PBH} \sim 10^{14}$\,g.

The antinuclei source spectrum can be written as 
\begin{equation}
    \mathcal{Q}_{\bar{p}(\bar{d})}\left(r,z, E\right) =   \int_{M_{\rm min}}^{M_{\rm max}} \; \d M\frac{g(r,z,M)}{M} \sum_{i = g, q, W, Z, h} \int_{m_i}^{\infty} \d E_i\left.\frac{\d^2 N_i (E_i,M)}{\d E_i \d t}\right|_{\rm HR} 
    \frac{\d N_{i \rightarrow \bar{p}(\bar{d})}}{\d E}\left(E, E_i\right)  \, ,
\label{eq:source_term_antinuclei}
\end{equation}
where the sum covers all quarks, gluons and bosons $i$, with mass $m_i$ and energy $E_i$, that are directly emitted by the PBH through Hawking radiation\footnote{Usually, in the PBH literature, particles directly produced in the evaporation process are dubbed as \textit{primary} (HR in this paper), while the subsequent hadronization/decay products as \textit{secondary}. This denomination might cause confusion in the rest of this paper, since CRs can also be of primary (directly produced from a Galactic source) or secondary (from spallation with gas atoms of the ISM) origin. In this paper, the term \textit{secondary} will refer to spallation products.}. The number of $\bar{p}$ or $\bar{d}$ with energy E that are produced by the parent particle $i$ is then obtained by convoluting the PBH evaporation emission with the fragmentation function, $\d N_{i \rightarrow \bar{p}(\bar{d})}\left(E, E_i\right)/\d E$. 
Since the physics of evaporation (and the subsequent hadronization) does not depend on the PBH numerical density, the source term can be factorized as~\cite{Barrau:2001ev}
\begin{equation}
    \mathcal{Q}_{\bar{p}(\bar{d})}\left(r,z, E\right) = q_{\bar{p}(\bar{d})}\left(E\right) \tilde{\rho}\left(r,z\right) \, ,
\end{equation}
where the energy-dependent part is
\begin{equation}
    q_{\bar{p}(\bar{d})}\left(E\right) =   \int_{M_{\rm min}}^{M_{\rm max}} \; \d M \; \frac{\d^2 N_{\bar{p}(\bar{d})}\left(E, M\right)}{\d E\d t} \frac{g_\odot(M)}{M} \,.
\label{eq:source_term_general}
\end{equation}
and  $g_\odot$ is the evolved lognormal PBH distribution at the position of the Sun.
The term $\d^2 N_{\bar{p}(\bar{d})}\left(E, M\right)/\d E\d t$ can be easily read in the right-hand side of Eq.~\eqref{eq:source_term_antinuclei}. 
The term with spatial dependence is the PBH density normalized to the local DM density $\tilde{\rho}= \rho_{\rm PBH}\left(r,z\right)/\rho_\odot$.

In our analysis, we rely on the public code {\tt BlackHawk v2.3}~\cite{Arbey:2019mbc,Arbey:2021mbl} to evaluate the HR spectra. To compute the hadronization contribution, we take the fragmentation functions from the tables provided in the \texttt{CosmiXs}~\cite{Arina:2023eic, DiMauro:2024kml} repository for an annihilating DM candidate\footnote{Note that \texttt{CosmiXs} always assumes the production of a particle-antiparticle jet. Therefore, we divide the original $\d N/\d E$ by a factor 2, to account for a single jet (we assume the second one not to escape the PBH), while the centre-of-mass energy is taken to be twice that of a single jet ~\cite{Herms:2016vop}. In doing so, we neglect any possible subdominant effects from cross-correlations among the two jets. We also assume that both quarks and antiquarks have the same probability of being reabsorbed into the PBH or escaping (and then to hadronize).}, fixing $m_{\rm DM} = E_i$.
The calculations performed in \texttt{CosmiXs} are based on the \texttt{Pythia} shower model version 8.309~\cite{Bierlich:2022pfr}, with \texttt{Vincia} shower plugin as default. The latter improves previous  \texttt{Pythia} versions on several aspects~\cite{Arina:2023eic}, focusing on particle DM annihilation spectra, which are similar to the PBH evaporation ones. 
The \texttt{CosmiXs} spectra (for antideuteron production \cite{DiMauro:2024kml}) include a coalascence model based on the Wigner formalism and calibrated on collider data, which reduces the uncertainties (due to nuclear fusion) to a few percent. 
The convolution between the HR spectrum and the fragmentation function is eventually performed with a custom \texttt{Python} code. 

\begin{figure}[t]
    \centering
\includegraphics[width=0.8\textwidth]{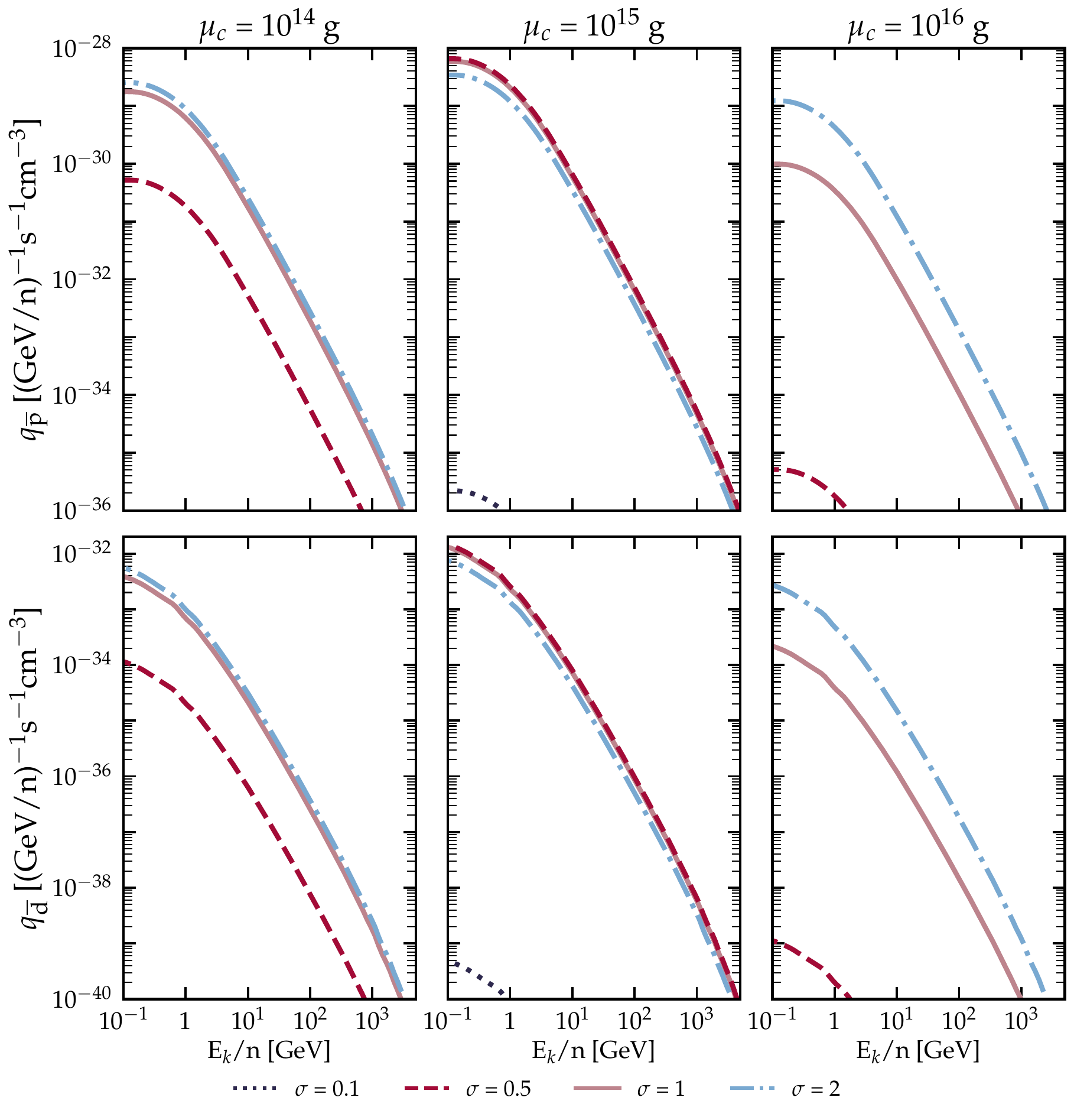}
\caption{The antiproton (top panels) and antideuteron (bottom panels) source spectra from PBH evaporation, as a function of the kinetic energy $E_k$ per nucleon. The spectra have been evaluated for PBHs with a lognormal mass distribution and assuming $\rho_{\rm PBH}=10^{-32}$\,g\,cm${^{-3}}=5.6\times10^{-9}$\,GeV\,cm${^{-3}}$, for different values of the critical masses $\mu_c$ and standard deviation $\sigma$, see the legend.}
    \label{fig:source_spectra}
\end{figure}
We show in Fig.~\ref{fig:source_spectra} the antiproton (top panels) and antideuteron (bottom panels) source spectra, Eq.~\eqref{eq:source_term_general}, as a function of the kinetic energy $E_k$ per nucleon\footnote{The total energy $E$ is related to the kinetic energy through $E = E_k + m_{\bar{p}(\bar{d})}$.}, assuming the lognormal mass distribution of PBHs in Eq.~\eqref{eq:lognormal}  evolved to the age of the Universe, with different critical masses and 
widths. Each panel corresponds to a different $\mu_c$, and displays the source spectra for $\sigma$ = 0.1 (dotted, blue), $\sigma$ = 0.5 (dashed, amaranth), $\sigma$ = 1.0 (solid, pink), $\sigma$ = 2.0 (dash-dotted, light blue).
In the computation of the source terms, we fix $M_{\rm min} = 10^{9}$\,g and $M_{\rm max} = 10^{18}$\,g, noting that $g(M[t_U])/M[t_U]$ is extremely suppressed below and above these masses (see Fig.~\ref{fig:dndM}).

As anticipated, most of the antinuclei fluxes originate from the population of PBHs that are evaporating now, hence for $\mu_c \simeq M_{\rm in}^*$. 
Below that mass, the PBHs that would be hot enough to produce antiparticles have already completely evaporated; above $M_{\rm in}^*$, the PBH distribution is not significantly affected by the time evolution, hence PBHs remain too cold to lead to a meaningful production for both antiprotons and antideuterons. 
As a consequence, PBHs with narrow mass distributions ($\sigma = 0.1, 0.5$) contribute significantly to antimatter fluxes only when $\mu_c \simeq M_{\rm in}^*$. Broader distributions can instead account for PBHs that are currently evaporating, even though they are peaked at smaller or larger $\mu_c$, translating into non-negligible $\bar{p}$ and $\bar{d}$ fluxes.
On the other hand, we do not display the source spectra of antimatter produced by PBHs with a monochromatic mass distribution, since their contribution would only be relevant for $M_{\rm mc} = M_{\rm in}^*$: only in the last instants of the evaporation process, the PBHs would become hot enough for the $\bar{p}$ and $\bar{d}$ production to be kinematically open. The monochromatic case can be seen as the asymptotic limit of a lognormal distribution with $\sigma \ll 0.1$. 
Note also that, in general, antideuteron source spectra are $\sim \mathcal{O} (10^{-4})$ smaller than antiproton ones, since their formation depends on the probabilities for an antiproton to be formed and then to coalesce with an antineutron. 
Our results are in general agreement with Ref.~\cite{Herms:2016vop}, although different fragmentation functions have been employed. 

An interesting feature of the computed source spectra is that they all share the same energy spectral shape, differing only by an overall normalization factor. We have verified that deviations from this behavior remain significantly below the percent level across all combinations of $\mu_{\rm{c}}$ and $\sigma$ that allow for antiproton production. This is due to the fact that the evolved mass distributions share a common shape proportional to $M^2$ for $M \lesssim 10^{13}$\,g, which is the mass range where PBH temperatures are high enough to produce antiprotons, as previously mentioned. This feature also holds for the power-law distribution, as well as for all mass distributions that extend to $M_{\rm in}=M_{\rm in}^*$ at $t = 0$, and therefore develop the low-mass tail due to evolution.

\section{Fluxes of antimatter from PBHs and astrophysical background}
\label{sec:fluxes}

Once the source spectrum for the evaporation of a PBH into an antinucleus is derived, it must be propagated to Earth and eventually compared to data. In this paper, we closely follow the calculations presented, for instance, in Ref.~\cite{Calore:2022stf}. 

CR transport in the Galaxy is described by a transport equation \cite{1990acr..book.....B,2002cra..book.....S,Strong:2007nh}. The steady-state equation for the differential density $n$ of Galactic CR antinuclei is taken in a cylindrical symmetry (coordinates $r,z$) with a diffusion zone of radius $R_G=20$\,kpc and variable half-height parameter $L$. In the thin disc approximation (half-height $h=100$\,pc), the transport equation reads

\begin{eqnarray}
    &-&\left[K\left(\frac{\partial^{2}}{\partial z^{2}} +
     \frac{1}{r}\frac{\partial}{\partial r}
     \left(r\frac{\partial}{\partial r} \right)\right)
     - V_{c} \frac{\partial}{\partial z}\right] n
     +2h\,\delta(z) \frac{\partial}{\partial E} \left[ b(E)\,n  - c(E)\, \frac{\partial n}{\partial E} \right]
     \nonumber\\
     &=& q^{\rm prim}(E) + 2h\delta(z) \Big[q^{\rm sec}(E)+q^{\rm ter}(E)\Big]
     - 2h\,\delta(z)\sum_{t\in \rm ISM} n_t v\,\sigma_{\rm inel}\;n\,. 
     \label{eq:2D_equation}
\end{eqnarray}
The following parameters and definitions hold:
\begin{itemize}
        \item {\em Transport parameters}: they include an isotropic spatial diffusion coefficient $K(R)$,  function of the rigidity, with a possible low- and high-rigidity break 
        \citep{2019PhRvD..99l3028G} (to which we refer for all details regarding the following equations including the definitions of the various parameters)
        \begin{equation}
          \label{eq:def_K}
          K(R) = {\beta^\eta} K_{0} \;
          {\left\{ 1 + \left( \frac{R}{R_{\rm l}} \right)^{\frac{\delta_{\rm l}-\delta}{s_{\rm l}}} \right\}^{s_{\rm l}}}
          {\left\{  \frac{R}{R_0=1\,{\rm GV}} \right\}^\delta}\,
          {\left\{  1 + \left( \frac{R}{R_{\rm h}} \right)^{\frac{\delta-\delta_{\rm h}}{s_{\rm h}}}
            \right\}^{-s_{\rm h}}}\,,
        \end{equation}
         a constant convective wind perpendicular to the disc with velocity $V_c$, and an effective Alfvén velocity $V_a$ mediating reacceleration via the momentum diffusion term 
         \begin{equation}
         K_{pp} =\frac{4}{3} V_a^2\beta^2 E^2 \frac{1}{\delta(4-\delta^2)(4-\delta)K(R)}\,.
         \end{equation}

        \item {\em First and second order energy redistribution terms}: they are respectively given by $b(E) = \left\langle\frac{dE}{dt}\right\rangle -E_k\left(\frac{2m+E_k}{m+E_k}\right) \frac{V_c}{3h}+(1\!+\!\beta^{2}) \frac{K_{pp}}{E}$ and $c(E) = \beta^{2} K_{pp}$, including ionization losses on neutral hydrogen and Coulomb losses in the ionized ISM, adiabatic losses, and reacceleration. We take $n_{\rm ISM}=1~$cm$^{-3}$ with 90\% H and 10\% He in number, and $\langle n_e\rangle=0.033$~cm$^{-3}$ and $T_e=10^4$~K.

        \item {\em Source and sink terms}: the source terms involve a {\em primary} source from PBH evaporation in the diffusive halo of the Galaxy, Eq.~(\ref{eq:source_term_antinuclei}), the {\em secondary} and {\em tertiary} source terms from standard nuclear interactions of Galactic CRs on the ISM (in the thin disc), written as
        \begin{eqnarray}
                q^{\rm sec}(E) &=& \int_{E'_{\rm th}}^{\infty} \d E'  \sum_{c \in \rm CRs}\,\left[\sum_{t\in \rm ISM} \left( n_t v' \frac{\d\sigma_{\rm prod}}{\d E}(E',E)\right) n^c(E')\right]\,,
        \label{eq:qsec}\\
                q^{\rm ter}(E) &=& \int_{E}^{\infty} \d E' \left[\sum_{t\in \rm ISM} \left( n_t v' \frac{\d \sigma_{\rm nar}(E',E)}{\d E}\right) n(E')\right] - \sum_{t\in \rm ISM} \Big( n_t v \sigma_{\rm ina} (E)\Big) n(E)\,,
                \label{eq:qter}
        \end{eqnarray}
        where we explicitly write the energy $E$ of the outgoing antinucleus, and where primed quantities in the integrals, $E'$ and $v'$, correspond to the incoming energy and velocity of the CR nucleus involved in the production of the antinucleus. The various cross-sections are related to the differential antiproton production, differential and total non-annihilating rescattering, and also in the last term of Eq.~(\ref{eq:2D_equation}), the total inelastic cross-section (sink term). The ingredients  of these secondary and tertiary terms are taken according to Ref.~\cite{Boudaud:2019efq}. An overview of the current status of these cross-section and their uncertainties can also be found in Ref.~\cite{Maurin:2025gsz}.
\end{itemize}

The semi-analytical solutions of the transport equation in cylindrical geometry---relying on Fourier-Bessel expansions---can be found in Refs.~\cite{2001ApJ...563..172D,Barrau:2001ev}. These solutions are implemented and available in the \usine{} code \cite{2020CoPhC.24706942M}, which we use in our analysis. As detailed in Ref.~\cite{Calore:2022stf}, for the primary contribution, we ensure the precision of the calculation by taking enough Bessel orders in the expansion and enforcing a smoothing of the strongly peaked DM distribution towards the Galactic center. In our main analysis, following Ref.~\cite{Calore:2022stf}, we use as a benchmark a NFW profile~\cite{Navarro:1995iw}  with a scale radius $r_s=19.6$\,kpc (to follow Ref.~\cite{McMillan_2016}) and a local DM density $\rho_\odot = 0.385$\,GeV\,cm$^{-3}$ (the latter is in the range 0.3--0.6\,GeV\,cm$^{-3}$ \cite{2021RPPh...84j4901D}), with $R_\odot = 8.20$ kpc~\cite{2019AandA...625L..10G}.
The differential density of antiprotons (primary or secondary) is converted into a differential flux (assuming isotropy), $\psi^{\bar{p}} = v/(4\pi) \times n^{\bar{p}}$, and we modulate the interstellar (IS) fluxes into top-of-atmosphere (TOA) fluxes using the force-field approximation for solar modulation \cite{GleesonEtAl1968a,Fisk1971}.

Both the primary and secondary $\bar{p}$ flux calculations depend on the values (or rather combinations) of the transport parameters, namely $K(R)/L$, $V_c$, and $V_a$, and on the diffusive halo size $L$. The former are obtained from the analysis of Li/C, Be/B, and B/C data, and the latter from the $^{10}$Be/$^9$Be ratio (using the radioactive clock $^{10}$Be, decaying into $^{10}$B with a lifetime of $1.387$~Myr).
Following \cite{2019AandA...625L..10G}, the generic form employed in Eq.~(\ref{eq:def_K}) is split in three benchmark configurations with varying numbers of transport parameters: (i) \BIG{} (parameters $K_0$, $\delta$, $R_l$, $\delta_l$, $s_l$, $V_c$, $V_a$, with $\eta=1$), \SLIM{} without convection and reacceleration (parameters $K_0$, $\delta$, and $R_l$, $\delta_l$, $s_l$), and \QUAINT{} (parameters $K_0$, $\delta$, $\eta$, and $V_c$, $V_a$); the remaining parameters are fixed. Following Refs.~\cite{Boudaud:2019efq,Calore:2022stf}, the fiducial model for all our analyses is \BIG{}.

Since, as discussed above, all the computed primary source spectra from PBHs share the same spectral shape and differ only by a normalization factor, this property is preserved in the propagated fluxes of primary antinuclei.

\section{Bounds from antiproton data }
\label{sec:pbar}
Data on CR antiprotons have been collected by AMS-02 
 with unprecedented precision and over an extended rigidity ranging from 1 to 525\,GV~\cite{AMS:2016oqu,AMS:2021nhj}.  These data can be used to set stringent bounds on the presence of antiprotons due to PBH evaporation.

\subsection{Statistical analysis and data}
\label{sec:statistical}
Following Ref.~\cite{Calore:2022stf}, the log-likelihood function of the analysis is taken to be
\begin{equation}
\label{likelihood_simple}
- 2 \ln {\cal L}(L,\mu) = \sum_{i,j} x_i({\cal C}^{-1})_{ij}x_j + \left\{ \frac{\log L - \log \hat{L}}{\sigma_{\log L}} \right\}^{2} \,,
\end{equation}
that is, the sum of the $\chi^2$ distance between the model and the data, and a nuisance parameter for the halo size $L$. The halo thickness $\hat{L}$ minimizes the fit on the nuclei fluxes ~\cite{Calore:2022stf}.  
In the first term, we have $x_i \equiv \psi_{i}^{\rm exp} - \psi_{i}^{\rm th}(L,\mu)$, the notation $\mu$ is generic to depict the PBH parameters (e.g., $\rho_{\rm PBH}$, $\mu_c$ and $\sigma$), the $i$ and $j$ indices run over the energy bins in the data vector, and the total covariance matrix ${\cal C}$ (see Ref~\cite{Boudaud:2019efq}) encodes existing energy correlations in experimental data (statistical and systematic uncertainties) and theoretical $q^{\rm sec}$ calculation (production and transport uncertainties). The primary flux being subdominant in $ \psi_{i}^{\rm th}$, it is assumed that the model covariance is fully encoded by the secondary flux one.
In principle, a fully consistent analysis should minimize over all transport and PBH parameters, considering simultaneously all secondary-to-primary, radioactive, and antiproton data. However, as argued in Ref.~\cite{Calore:2022stf}, a much faster and more resource-friendly approach is to use the above log-likelihood, which more directly includes the transport uncertainties (via the covariance matrix) and the halo-size uncertainties (via the penalty on $L$). The latter is centered on $\hat{L}$ with a symmetrized lognormal spread $\sigma_{\log L}$, whose values are taken from Table~3 (first half) of Ref~\cite{Weinrich:2020ftb} for the \BIG{}, \SLIM{} and \QUAINT{} configurations.
This halo size plays a very important role. Indeed, whereas the secondary component is mostly independent of combinations of the transport parameters, the primary component breaks this degeneracy, its flux roughly scaling with $L$ (see Refs.~\cite{Barrau:2001ev, Donato:2003xg, Genolini:2021doh}).

To go even faster, our analysis goes as follows: we pre-calculate with the USINE 
code \cite{2020CoPhC.24706942M} the IS primary flux on a grid of $L$ and $\mu_c$ values (at fixed $\sigma$ and for a reference $\rho_{\rm PBH}^{\rm ref}$), and re-use the IS secondary fluxes calculated on a grid of $L$ values in Ref.~\cite{Calore:2022stf}; we also re-use the covariance matrix ${\cal C}$ calculated in Ref.~\cite{Boudaud:2019efq}. Interpolation functions of these IS fluxes allow us, after modulating the IS into TOA fluxes, to calculate instantaneously combinations (see below) of the log-likelihood, Eq.~\eqref{likelihood_simple}, for any $L$ and $\mu_c$ values, but also for any $\rho_{\rm PBH}$ (by a mere rescaling of the primary flux by $\rho_{\rm PBH}/\rho_{\rm PBH}^{\rm ref}$). This calculation can be repeated for different discrete $\sigma$ values, transport configurations, and DM halo profiles.

In our analysis, we are mostly interested in establishing upper bounds on the PBH density, $\rho_{\rm PBH}$, at fixed $\mu_c$ and $\sigma$. Following Ref.~\cite{Calore:2022stf}, we rely on the likelihood ratio (LR),
\begin{equation}
{\rm LR}(\rho_{\rm PBH}) = - 2 \ln {\cal L}(L_{\rm min},\rho_{\rm PBH}) \, + \, 2 \ln {\cal L}(L',\rho_{\rm PBH}') \,,
\label{eq:upperlimit}
\end{equation}
with ${\cal L}$ taken from Eq.~\eqref{likelihood_simple}. In the above expression, the first term is calculated at $L_{\rm min}$ (halo size maximizing ${\cal L}$ for a given $\rho_{\rm PBH}$), and the second one at $(L',\rho_{\rm PBH}')$ maximizing ${\cal L}$. According to Wilks' theorem, the LR is distributed as a $\chi_\nu^2$ with $\nu=1$ degree of freedom (d.o.f) in our case (only $\rho_{\rm PBH}$ is varied). Following the standard convention in the literature, the upper limit (UL) is calculated from the 95\% confidence level (CL) on this distribution, which amounts to finding $\rho_{\rm PBH}^{\rm UL}$ for which the LR increases by 3.84.
We show in the next paragraph the best-fit fluxes obtained by directly maximizing Eq.~\eqref{likelihood_simple}, for illustration purposes only. Indeed, the secondary antiproton flux was found to be compatible with the AMS-02 data in Ref.~\cite{Boudaud:2019efq}, and a careful analysis in the context of DM annihilation, carried out in Ref.~\cite{Calore:2022stf}, confirmed that the presence of component other than secondaries was not statistically significant. This conclusion was reached using the antiprotons data released by the AMS-02 collaboration in 2016~\cite{AMS:2016oqu}, and was also found to hold for their larger dataset ($5.6\times 10^5$ vs. $3.5\times 10^5$ events) released in 2021~\cite{AMS:2021nhj}---we refer to Ref.~\cite{Calore:2022stf} for a thorough discussion of the possible new challenges and statistical complication associated with the interpretation of these higher precision data.

Below, we test both AMS-02 data sets,
assuming the \BIG{} propagation set-up as the reference, and an average solar modulation Force-Field (or Fisk) potential  $\langle\phi\rangle=731$ MV, as determined in Refs.~\cite{Boudaud:2019efq,Genolini:2021doh}. 
Uncertainties on $\langle\phi\rangle$ are already captured in the covariance matrix used in Eq.~\eqref{likelihood_simple}. Indeed, the covariance matrix encodes, among others, the uncertainties and correlations from the modeling. The latter include the transport parameters uncertainties, which themselves include solar modulation uncertainties $\sigma_{\langle\phi\rangle}=100$\,MeV (via a nuisance parameter on $\langle\phi\rangle$, see Refs.~\cite{Weinrich:2020cmw, Weinrich:2020ftb}). As a result, we do not need to account for an extra $\sigma_{\langle\phi\rangle_{\bar{\rm p}}}$, provided nuclei and antinuclei are similarly modulated.
Actually, more realistic solar modulation models, either effective~\cite{2019PhRvL.123y1104K,2025ApJ...982..103Z}, or fully accounting for a drift effect \cite{Potgieter2013}, involve a charge-sign dependence. However, the number of parameters in these models rapidly grows and is still disputed. Thanks to the recently released AMS-02 time-dependent antiproton data~\cite{2025PhRvL.134e1002A}, the exact impact of the drift and possible hysteresis effects are being investigated (see, e.g., Refs.~\cite{2023ApJ...947...72A,2025arXiv250314025T}). Given that the Force-Field approximation gives an overall consistent picture of the secondary-to-primary and antiproton data (see Ref.~\cite{Boudaud:2019efq}), we stick to it. The usage of more complex modulation models would slightly change the uncertainty band on the modeled antiproton total flux, 
consequently shifting (upwards or downwards) the upper limits derived on $\rho_{\rm PBH}$, but would leave our conclusions unchanged.

\subsection{Results for the $\rho_{\rm PBH}$ upper bounds}
\begin{figure}[t]
    \centering
   \includegraphics[width=\textwidth]{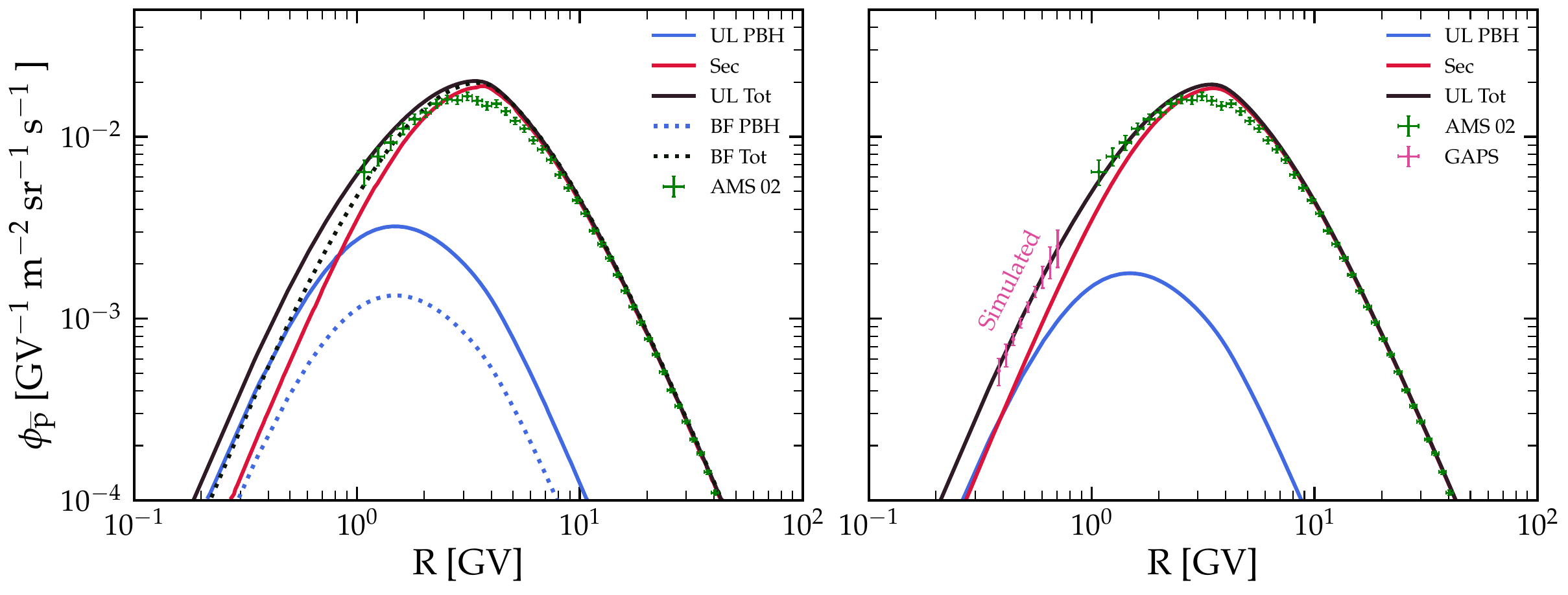}
\caption{Fluxes of top-of-atmosphere antiprotons: secondary flux (solid red line), primary flux from PBH evaporation from a population with a lognormal mass distribution $\mu_c = 10^{14}$\,g and $\sigma = 1$ (blue lines), and total flux (black lines). 
Left panel: the blue solid curve is the $95 \%$ CL UL on the PBH flux, obtained from the LR analysis in Eq.~\eqref{eq:upperlimit}; for illustration purpose, the dashed lines show the PBH (and total) fluxes fitting best the data. Green symbols are from AMS-02~\cite{AMS:2021nhj}.
Right panel: $95 \%$ CL UL derived from the combination of AMS-02 data and GAPS (simulated) data~\cite{GAPS:2022ncd} (pink symbols).}
\label{fig:flux_antiproton}
\end{figure}
As a first analysis, we minimize Eq.~\eqref{likelihood_simple} over the halo size and PBH parameters and compare our predictions with AMS-02 data \cite{AMS:2016oqu}, including a secondary component summed to the one from the PBH source.
We notice that the secondary component, as discussed in \cite{Calore:2022stf}, is not really fitted but is instead predicted based on previous analyses of the AMS-02 nuclear components. The data in \cite{AMS:2016oqu} are well explained  by a secondary component. The addition of a PBH flux does not improve the fit, since it must be very tiny not to overshoot the lowest-energy bins (the second one specifically). 

We show in Fig.~\ref{fig:flux_antiproton} (left panel) 
 the TOA antiproton flux produced by secondary processes and by the evaporation of Galactic PBHs with a lognormal mass distribution with $\mu_c = 10^{14}$ g and $\sigma =1$. The fit is performed on the AMS-02 data given in \cite{AMS:2021nhj}. 
The antiproton flux from PBHs is reported for both the best fit, added for illustrative purposes only (dotted, blue) and the UL as defined in Eq.~\eqref{eq:upperlimit} (solid, blue) at $95 \%$ CL, together with the total flux (solid, black) obtained by a sum to the secondaries (solid, red). 
The best-fit value is found for $\rho_{\rm{PBH}} = 3.7 \times 10^{-11} \, \rm{GeV/cm^3}$, while the UL is set to $\rho_{\rm{PBH}}= 1.9 \times 10^{-10} \, \rm{GeV/cm^3}$. 
The PBH flux shows a maximum around 1-2 GeV, slightly shifted to lower energies with respect to the secondary one. We interpret the position of this peak as a non trivial combination of the Hawking radiation convolved over the PBH mass distribution, the hadronization spectra of elementary particles, propagation in the Galaxy, and, finally, the effect of solar modulation, 
which alters the IS spectrum and prevents the lowest CRs to reach the Earth.  
This feature persists when changing the values of $\mu_c$ and $\sigma$, always imposing the fit to the AMS-02 data, as PBH antiproton fluxes share the same spectral shape and hence the value of $\rho_{\rm{PBH}}$ is merely rescaled when changing $\mu_c$ and $\sigma$.
This is due to the fact that the evolution of the mass spectrum follows the same $M^2$ power law below about $5\times 10^{14}$ g, which is the relevant range for the antiproton production. The scaling in the source spectrum is naturally maintained after Galactic propagation.  
 We stress that, contrary to other publications where error bands on the fluxes are shown to account for various uncertainties (propagation, production cross-sections, or solar modulation), our LR analysis (and derived UL) directly accounts for all of these uncertainties (see Sec.~\ref{sec:statistical}). Hence, a single UL curve is shown for all our analyses, as will also be the case for the constraints on the PBH density derived below.

We have also assessed the GAPS potential by analyzing their projected sensitivity to CR antiprotons, as reported in Ref.~\cite{GAPS:2022ncd} (see their Fig.~9), with the statistics expected from three 35-day flights. These TOA data would have the originality to be taken at very low energies, in the range 0.07--0.21\,GeV. 
In the right panel of Fig.~\ref{fig:flux_antiproton}, we report the PBH antiproton flux for the $95 \%$ CL ULs on the $\rho_{\rm PBH}$ derived by combining the AMS-02 data with the simulated GAPS ones. For simplicity, we have assumed the same modulation level for both data sets, i.e. $\langle\phi\rangle=731$\,MV,
also consistent with the solar modulation conditions  considered in~\cite{GAPS:2022ncd}.
The inclusion of GAPS data very nicely complement the AMS-02 ones towards the low-energy side of the PBH flux peak, i.e. close to where the constraining power should be maximal \cite{Aramaki:2014oda}.

\begin{figure}[t]
    \centering
\includegraphics[width=0.55\textwidth]{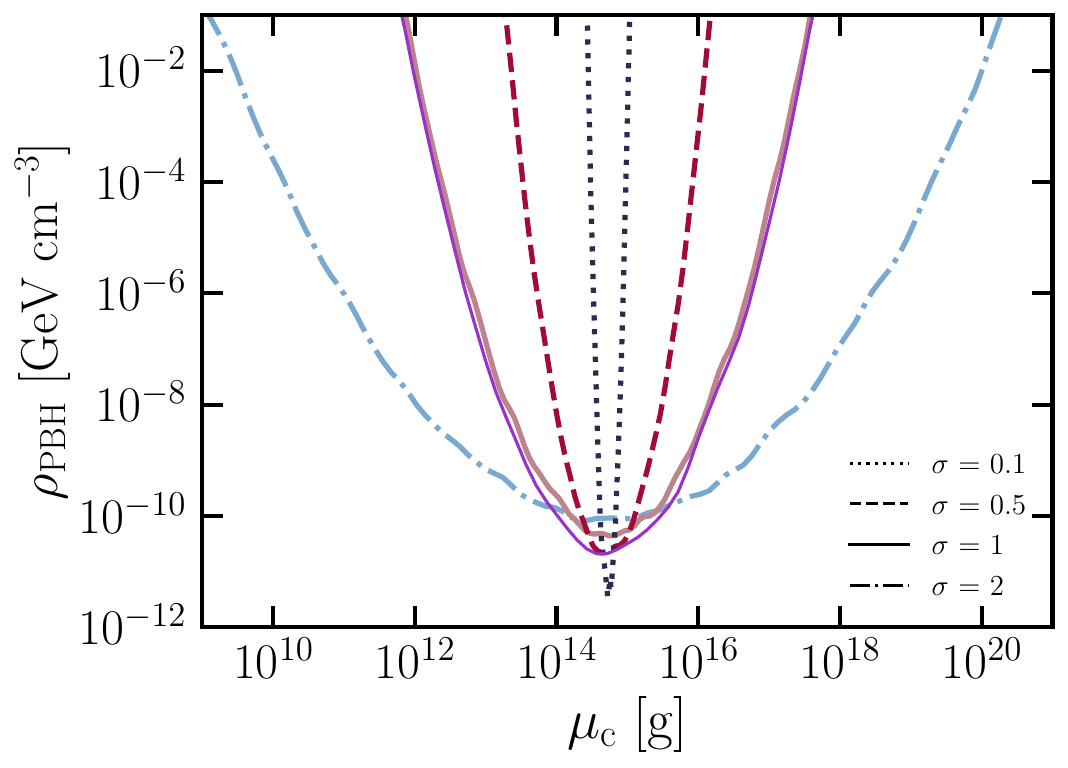}
\caption{Upper bounds for $\rho_{\rm{PBH}}$ as a function of $\mu_{c}$, for $\sigma=$ 0.1 (dotted), 0.5 (dashed), 1 (solid) and 2 (dot-dashed), after a fit on AMS-02 data \cite{AMS:2021nhj}. 
The thin solid purple line denotes the future sensitivity obtained from a combined fit of AMS-02 and the simulated GAPS data \cite{GAPS:2022ncd},  for $\sigma = 1$.}
    \label{fig:density_bounds}
\end{figure}

The upper bounds obtained for $\rho_{\rm{PBH}}$ after fitting the AMS-02 antiproton data \cite{AMS:2021nhj} are shown in Fig. \ref{fig:density_bounds} as a function of $\mu_c$, for different values of the lognormal parameter $\sigma$. 
The results strongly depend on the width of the lognormal distribution: they get stronger while $\sigma$ spreads from 0.1 to~2. 
The maximum allowed $\rho_{\rm{PBH}}$ for $\mu_{\rm c} \simeq 5\times 10^{14}$\,g is about 
$3.5 \times 10^{-12}$, $2.2 \times 10^{-11}$, $4.4 \times 10^{-11}$ and $8.1 \times 10^{-11}$\,GeV\,cm$^{-3}$, for $\sigma=0.1$, $0.5$, $1$ and $2$, respectively.
We have verified that changing the propagation setup from \BIG{} to \SLIM{} does not imply a perceptible modification in the upper bounds in Fig.~\ref{fig:density_bounds}, while the \QUAINT{} model requires a $\rho_{\rm{PBH}}$ a factor two smaller. 
A further check on the DM density distributions has confirmed expectations that using a contracted NFW or a cored profile leaves the results unchanged at the percent level\footnote{A larger effect is observed for annihilating DM (e.g., \cite{Calore:2022stf}), because the corresponding source terms involve the squared density profile.}.

Looking at the available and future datasets, we note that the fit to the older AMS-02 data \cite{AMS:2016oqu} would provide bounds that are about 50\% lower,  mostly related to the large fluctuations between the most constraining low-energy data points.
Conversely, the addition of low-energy GAPS data \cite{GAPS:2022ncd} for three hypothetical 35-day flights would increase the upper bounds by a factor~2, as shown in the figure for the case $\sigma=1$. 
The BESS Polar-II data antiproton data \cite{Bess_polar_ii_antiprotons_data} have not been considered here since their large errors would not improve the exclusion power brought by AMS-02 data.

Although better CR data are always desired, we stress that the derived UL could be improved thanks to new nuclear data. Indeed, as demonstrated in Refs.~\cite{Donato:2017ywo, Korsmeier:2018gcy,Boudaud:2019efq}, the uncertainty due to nuclear processes can reach 20\%. Therefore, a step forward in the evaluation of the room left to exotic physics, 
including the production from PBHs, addresses new measurements of the production cross sections on H and He targets at accelerators, as discussed extensively in~\cite{Maurin:2025gsz}.

\begin{figure}[t]
    \centering
   \includegraphics[width=0.495\textwidth]{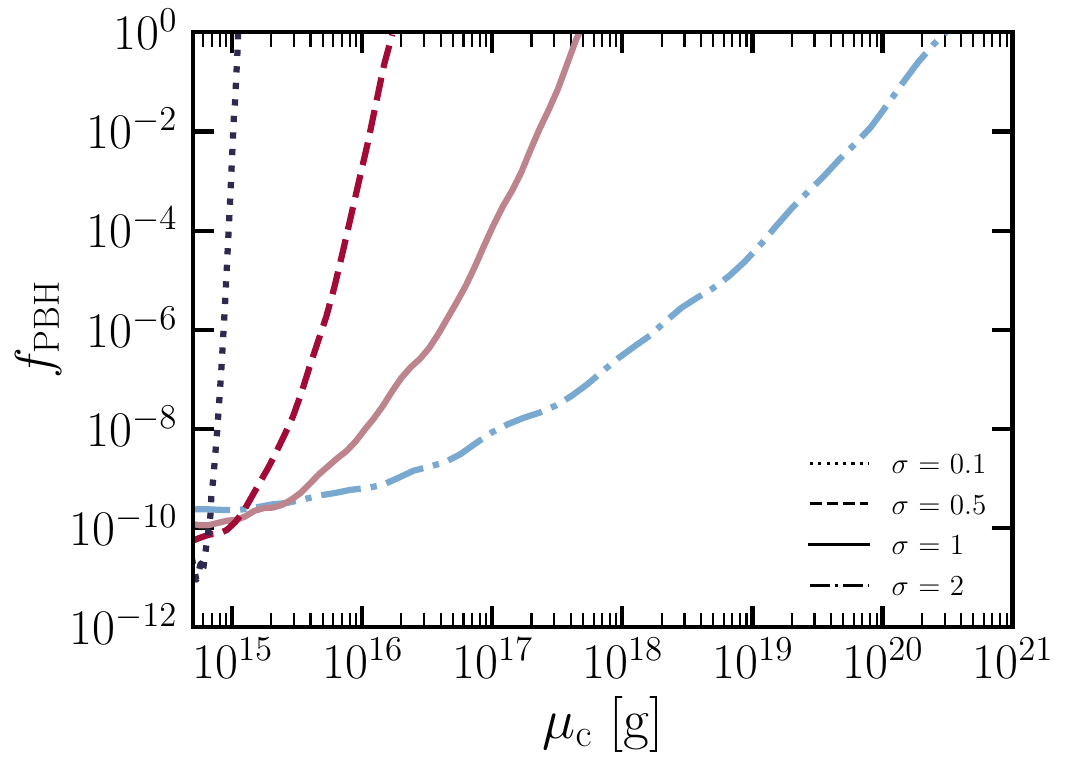}
      \includegraphics[width=0.495\textwidth]{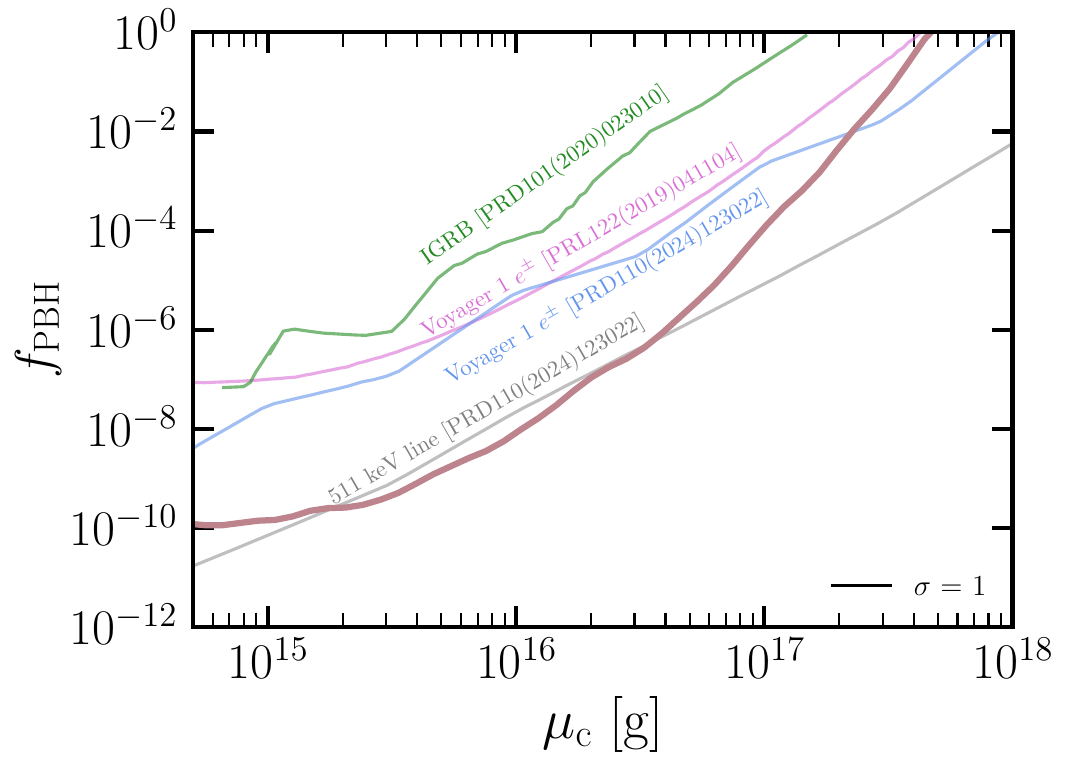}
\caption{Antiproton bounds on the fraction $f_{\rm PBH}$ of the PBH to the local DM density, $\rho_{\rm PBH}/\rho_\odot$, as a function of $\mu_{c}$.  
Left panel: $95 \%$ CL exclusion limits based on AMS-02 data \cite{AMS:2021nhj} for  a lognormal PBH mass distribution with width $\sigma=$ 0.1 (dotted), 0.5 (dashed), 1 (solid) and 2 (dot-dashed). 
Right panel: our result (red thick line) for $\sigma = 1$, compared with other bounds from different messengers, Voyager-1 $e^\pm$ from Refs.~\cite{Boudaud:2018hqb} and \cite{DelaTorreLuque:2024qms}, $\gamma$ rays in the IGRB~\cite{Arbey:2019vqx}, and the 511 keV line~\cite{DelaTorreLuque:2024qms}. 
}  \label{fig:f_bounds}
\end{figure}

We further translate the upper bounds obtained from the fit of AMS-02 antiproton data into the maximum fraction $f_{\rm PBH} = \rho_{\rm PBH}/\rho_{\odot}$ of local DM that can be composed of PBHs, where we recall we take $\rho_{\odot}= 0.385$\,GeV\,cm$^{-3}$. 
We focus on $\mu_{c}$ values greater than $5\times 10^{14}$\,g, reminding that PBHs with masses smaller than $M^* = 5 \times 10^{14}$\,g would have been completely evaporated by now, and therefore could not be viable candidates for DM. 
We show in Fig.~\ref{fig:f_bounds} (left panel) the $95 \%$ CL exclusion limits on $f_{\rm PBH}$ as a function of $\mu_c$ for different values of the lognormal width parameter $\sigma$. 
The strongest bound is obtained for $\sigma = 0.1$ and reads $f_{\rm PBH} < 10^{-11}$, at $\mu_c = M^*$. Broader mass distributions, despite leading to less stringent bounds than a narrow distribution, allow to probe heavier critical masses and the so-called \textit{asteroid-mass} region. Cases for which the DM could be entirely due to PBHs are practically excluded by AMS-02 antiproton data. 
To put our results into context, we present in Fig.~\ref{fig:f_bounds} (right panel) our bound for the case $\sigma = 1$ and compare it with other existing constraints in the literature, obtained for the same lognormal distribution width.  
In particular, we depict limits from: the isotropic gamma-ray background (IGRB, light green,~\cite{Arbey:2019vqx}); two analyses of low-energy IS electrons and positrons measured by the Voyager 1 spacecraft (light pink,~\cite{Boudaud:2018hqb}, and light blue,~\cite{DelaTorreLuque:2024qms}); and the diffuse 511 keV line measured by INTEGRAL/SPI (light gray,~\cite{DelaTorreLuque:2024qms}). 
The state-of-the-art upper bounds are all obtained with cosmic messengers different from CR antiprotons. Therefore, a direct comparison is not possible and would not be correct. Moreover, different models are assumed for the propagation of CR electrons and positrons, for the statistical analysis, and for the computation of the backgrounds to the PBH signals. 
However, a visual comparison helps to estimate the power of AMS-02 antiproton data, which turns out to be very competitive. 
Our bounds are stronger than the ones obtained with CRs, namely from Voyager $e^\pm$, by at least two orders of magnitude. 
For $\mu_{\rm c} \lesssim 5\times 10^{15}$ g,  they compare or slightly improve the results 
obtained from the diffuse 511 keV line measured by INTEGRAL/SPI. 

As previously discussed, the largest fluxes of antiprotons arise from the last instants of life of PBHs that are now evaporating. We can therefore connect the upper bounds previously obtained on the local PBH density to the rate of \textit{exploding} PBHs, as shown in~\cite{Maki:1995pa}. We estimate the local PBH explosion rate as~ \cite{Boluna:2023jlo,Maki:1995pa}
\begin{equation}
    \dot{n}_{\rm PBH} = \frac{g(r,z,M_{\rm in} = M^*_{\rm in})}{3t_U} \,,
\end{equation}
where $t_U$ is the age of the Universe. By imposing the $95 \%$ CL upper limit 
$\rho_{\rm PBH}|_{\mu_c = M^*} = 3.5\times \rm 10^{-12} ~GeV~ cm^{-3}$ for $\sigma=0.1$,
we infer the following limit
\begin{equation}
\label{eq:bound_expl}
    \dot{n}_{\rm PBH} \lesssim 7\times \rm 10^{-5}\,pc^{-3}\,yr^{-1} \,,
\end{equation}
applying to the lognormal distribution.
Note that this bound improves upon the previous one~\cite{Maki:1995pa} by more than two orders of magnitude, thanks to the increased precision of the data and the differences in the analysis approach. However, let us note that the translation of bounds into an explosion rate is strongly dependent on the observation time of the experiment, and hence the interval of explosion that is considered, as well as the chosen PBH mass range and distribution. Therefore, care should be taken when comparing different bounds~\cite{Barrau:2001ev}. All in all, the limit that we obtain in Eq.~\eqref{eq:bound_expl} under the above-mentioned assumptions is several orders of magnitude stronger than the one obtained from other messengers. For the sake of comparison, facilities searching for gamma-ray bursts in our Galaxy currently probe a much larger rate. Among them, the H.E.S.S array of imaging atmospheric Cherenkov telescopes~\cite{HESS:2023zzd} sets the strongest direct limit on the rate of exploding PBHs: $\dot n_\mathrm{PBH} < 2000\,\mathrm{pc}^{-3}\,\mathrm{yr}^{-1}$ for a burst interval of 120 seconds, at the $95 \%$ CL. 

Summarizing, we have performed an analysis of antiproton CR which improved the state-of-the-art under several aspects. As well as i) the adoption of a log-normal distribution for the differential PBH number density, our analysis relies on 
ii) the AMS-02 data on antiproton data, including the most recent data release; 
iii) an up-to-date propagation set up, which is tuned on the most recent CR nuclei data; 
iv) a sophisticated statistical analysis, which includes a covariance matrix encoding existing energy correlations in experimental data and theoretical calculations; 
v) the treatment of the Galactic halo thickness $L$ as a nuisance parameter, which let us obtain reliable upper limits of the PBH fraction, 
vi) the new CosmiXs antiproton fragmentation functions.
All these innovative elements have lead to new constraints on the Galactic PBH density, which are not necessarily stronger than the few existing ones (which indeed are almost impossible to directly compare), but are definitely very solid, being rooted on very precise experimental data and accurate theoretical models.

\section{Perspectives with antideuterons measurements }
\label{sec:dbar}
Similarly to antiprotons, heavier antinuclei  can be produced in space by secondary reactions and, if any, by primary sources such as evaporating PBHs, or DM annihilation/decay. 
It has been demonstrated that the secondary production of antideuteron and antihelium is strongly suppressed at energies below a few GeV
\cite{Donato:1999gy,Barrau:2002mc,Donato:2008yx,Korsmeier:2017xzj,Cirelli:2014qia,Carlson:2014ssa}, thus letting open a discovery window for exotic sources in the Galactic halo. 
The major difference with antiproton production lies in the fusion of the antinucleons, which is prevented for  energetic antinucleons.

\begin{figure}[t]
    \centering
   \includegraphics[width=0.7\textwidth]{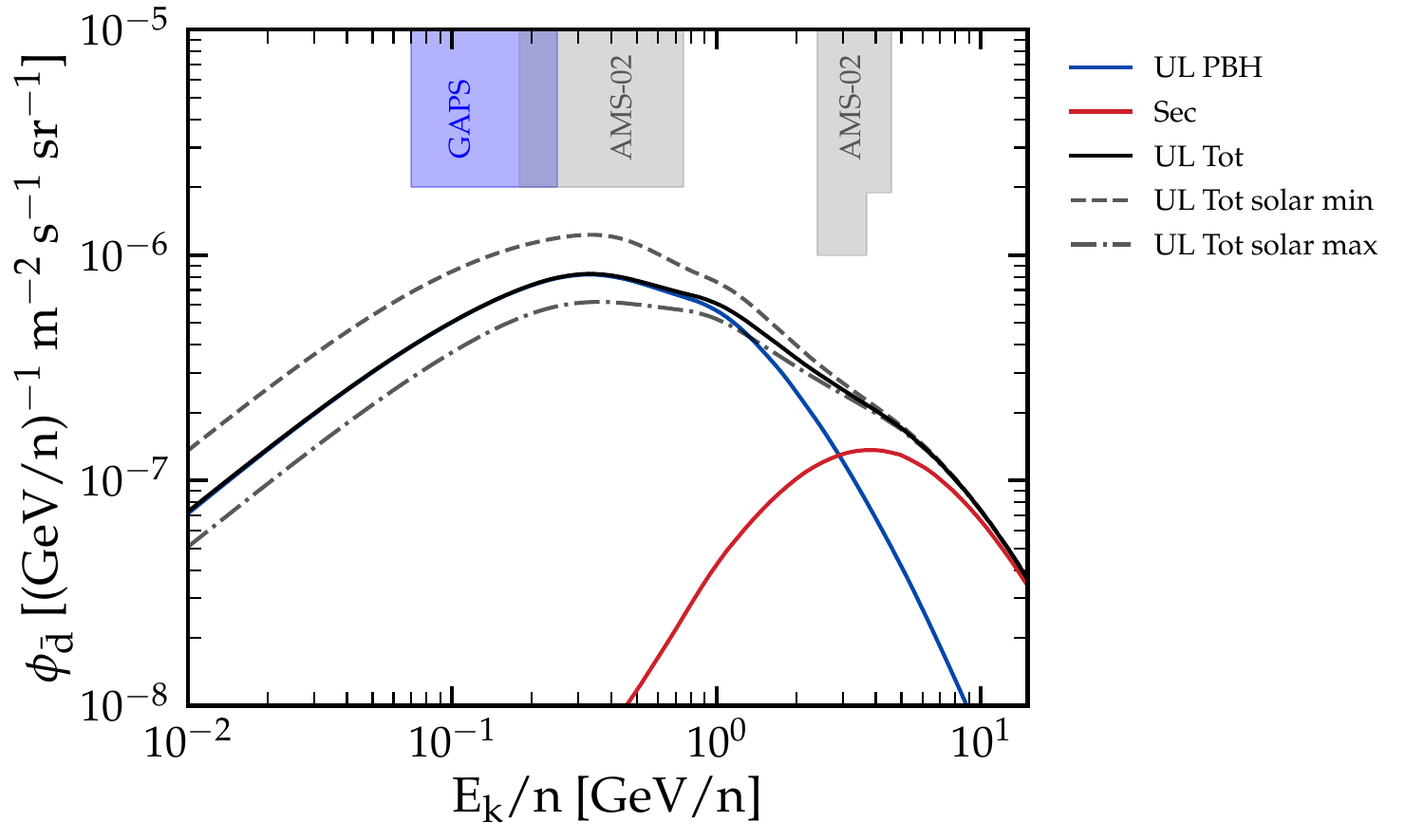}
\caption{Fluxes of TOA antideuterons as due to secondary processes (solid red line) and to the evaporation of a population of Galactic PBHs with a lognormal mass distribution with $\mu_c = 10^{15}$\,g and $\sigma = 1$ (solid blue line). The value of $\rho_{\rm{PBH}}$ is set to $5.2 \times 10^{-11}$\,GeV\,cm$^{-3}$, corresponding to the $95 \%$ CL UL derived from the antiproton analysis for this set of PBH parameters. 
The solid black line is the sum of the two components for a solar modulation potential 
$\langle \phi \rangle = 731$\,MV, while the dashed (resp. dot-dashed) line is for $\langle\phi\rangle =$ 500 (resp.~900)\,MV. 
The blue and gray shaded regions indicate the expected antideuteron sensitivities of GAPS \cite{Aramaki_2016} and AMS-02 \cite{ARAMAKI20161} experiments, respectively.
 }
    \label{fig:dbar_fluxes}
\end{figure}
We compute the CR antideuteron fluxes from PBHs and secondary origin following the same \BIG~propagation scheme and parameters as for antiprotons.
Our main results are displayed in Fig. \ref{fig:dbar_fluxes}. The flux for TOA antideuteron of secondary origin (solid red line) is shown along with the one for PBHs for  $\mu_c = 10^{15} \, \rm{g}$, $\sigma = 1$ and $\rho_{\rm{PBH}}= 5.2 \times 10^{-11}\,\rm{GeV\,cm^{-3}}$ (solid blue line), which is the upper limit allowed by antiproton data (see Sec. \ref{sec:pbar}). The solar modulation potential is set to $\langle \phi \rangle = 731$\,MV. 
The total flux given by the sum of the two components, is shown by the black solid line. 
The dashed (dot-dashed) line corresponds to  $\langle\phi\rangle =$ 500 (900)\,MV, indicative  for solar minimum (maximum) conditions, respectively.  
We also report the experimental sensitivities for GAPS \cite{Aramaki_2016} and AMS-02 \cite{ARAMAKI20161}. We show the flux for a single choice of $\mu_{c}$ and $\sigma$, but we have verified that the results remain unchanged for all other values considered. This is because all computed antideuteron fluxes share the same spectral shape, and the constraining procedure on AMS-02 antiproton data ultimately corresponds to a rescaling of $\rho_{\rm{PBH}}$, which is reflected in the normalization of the antideuteron fluxes.

The maximal antideuteron flux from PBHs allowed by AMS-02 data on antiproton is increasingly higher than the secondary one for energies below 3 GeV. 
If one or more antideuterons would be measured by AMS-02 or by GAPS, it would be clearly a signal of new physics, hardly explainable with PBH emission only. 
As already noted for antiprotons, uncertainties inherent to the propagation parameters are automatically accounted for in the likelihood minimization procedure (see Sec.~\ref{sec:statistical}). The fluxes plotted in Fig.~\ref{fig:dbar_fluxes} are therefore to be considered as maximal with respect to the transport model. 
Further uncertainties due to the fusion modeling of the antinuclei to for and antideuteron are very small, a few percent for kinetic energies relevant for AMS and GAPS, as estimated in \cite{DiMauro:2024kml}. 
Our results predict a maximal antideuteron flux which is significanly higher than what found in \cite{Herms:2016vop, Serksnyte:2022onw}, as it is found already for antiprotons. 
Possible reasons for these difference can be traced back to a different prediction for the secondary antiproton flux, to resorting to AMS-02 data in \cite{AMS:2016oqu} rather than in \cite{AMS:2021nhj} and, to a lesser extent, different propagation models. 
Finally, our model for coalescence into antideuterons is updated to \cite{DiMauro:2024kml}, even if the differences with the benchmark model in \cite{Herms:2016vop} are indeed small. 

The analysis that we have presented here could be easily extended to antihelium searches \cite{Carlson:2014ssa,Cirelli:2014qia,Herms:2016vop}. 
However, the antihelium-3 production is  suppressed by more than three orders of magnitude with respect to the antideuteron one. Given the results we have just derived on antideuteron fluxes,  the perspectives for searching antihelium from PBHs in CRs are far for being relevant.

\section{Conclusions}
\label{sec:conclusions}
The emission of particles through Hawking evaporation from PBHs provides an interesting opportunity to assess their role in constituting the DM halo of our Galaxy. Charged antiparticles can in principle be produced after the hadronization of quarks and gluons, and can be measured as GeV or sub-GeV CRs. 
Thanks to recent space-based experiments, data
from antiprotons have been collected on a wide energy extension and with very high precision. 
In the near future, the GAPS experiment will undertake
a series of Antarctic long-duration balloon flights, measuring the antiproton flux in the sub-GeV region, and improving the current antideuteron sensitivity by two orders of magnitude. 

In this paper, we revisited the CR antiproton and antideuterons signatures from PBH evaporation in the Galaxy. We improved previous calculations on many aspects, considering an extended lognormal PBH mass distribution, using state-of-the-art Galactic propagation parameters, and, more importantly, analyzing the most recent AMS-02 antiproton data within a sophisticated statistical framework. 
We observed a universal spectral behavior of the antinuclei fluxes, the normalization depending on a combination of $\mu_c$, $\sigma$ and $\rho_{\rm PBH}$. The direct consequence is that the solar modulated primary PBH fluxes peak at $\sim$~1--2\,GeV for antiprotons and at a few hundreds of MeV/n for antideuterons. CR data around and below these peaks provide the best energy range to set constraints on the PBH density in the Galaxy. 
We found that the AMS-02 data set strong upper bounds on the local PBH density. The maximum allowed $\rho_{\rm{PBH}}$ for $\mu_{\rm c} = M^* \simeq 5\times 10^{14}$\,g is about 
$3.5 \times 10^{-12}$, $2.2 \times 10^{-11}$, $4.4 \times 10^{-11}$ and $8.1 \times 10^{-11}$\,GeV\,cm$^{-3}$, for $\sigma=0.1$, $0.5$, $1$ and $2$, respectively.
The ULs follow a rough parabolic shape for $\rho_{\rm{PBH}}$ vs $\mu_{\rm c}$, which gets wider for increasing $\sigma$ values. Moving away from $5\times 10^{14}$\,g, higher $\sigma$ values dig into lower $\rho_{\rm{PBH}}$. 
Our statistical analysis automatically accounts for uncertainties due to propagation in the Galaxy, so that these numbers have to be considered as maximized also in these terms.  
We have also assessed the GAPS potential by analyzing their projected sensitivity to cosmic
antiproton spectrum, and the forthcoming data from this instrument should improve the above limits by a factor 2. 

When translating these constraints into the fraction of the local DM density, we found that the strongest bound is obtained for $\sigma = 0.1$, with $f_{\rm PBH} < 10^{-11}$ at $\mu_c = M^*$. Broader mass distributions allow to probe heavier critical masses and the so-called \textit{asteroid-mass} region, despite being less stringent than for a narrower distribution. Cases for which the DM could be entirely due to PBHs are practically excluded by AMS-02 antiproton data for broader mass distributions. 
A comparison with state-of-the-art bounds obtained with different cosmic messengers---although only indicative given several different assumptions---indicates that our results obtained with AMS-02 data are very competitive. 

 Although the antideuteron flux from PBHs can be orders of magnitude higher than the secondary one below a few GeV/n, its predicted intensity, when bound to AMS-02 antiproton data, is barely skims the sensitivity thresholds of the GAPS and AMS-02 experiments.
We confirm that if one or more antideuterons were to be measured by AMS-02 or by GAPS, they would clearly be a signal of new physics. 
However, as it has been proved in this work, they would hardly be entirely explainable by PBH emission.

\section*{Acknowledgments}
 We are indebted to J. Herms for sharing his Master thesis, to M. Di Mauro for valuable help with \texttt{CosmiXs}, and  to S. Gariazzo for technical computing assistance.  
 We are grateful to F. Giovacchini, I. Masina and  Y. F. Perez-Gonzalez for useful discussions.
F.D. and L.S. are supported by the 
Research grants {\sl The Dark Universe: A Synergic Multimessenger Approach}, No.~2017X7X85K, funded by the {\sc Miur} and  {\sl TAsP (Theoretical Astroparticle Physics)} funded by Istituto Nazionale di Fisica Nucleare (INFN).
V.D.R. and A.T. acknowledge financial support by the grant CIDEXG/2022/20 (from Generalitat Valenciana) and by the Spanish grants CNS2023-144124 (MCIN/AEI/10.13039/501100011033 and “Next Generation EU”/PRTR), PID2023-147306NB-I00, and CEX2023-001292-S (MCIU/AEI/10.13039/501100011033).
\bibliographystyle{JHEP}
\bibliography{main}

\providecommand{\href}[2]{#2}\begingroup\raggedright\begin{thebibliography}{100}

\bibitem{Hawking:1971ei}
S.~Hawking, \emph{{Gravitationally collapsed objects of very low mass}}, {\emph{Mon. Not. Roy. Astron. Soc.} {\bfseries 152} (1971) 75}.

\bibitem{Zeldovich:1967lct}
Y.B.~Zel'dovich and I.D.~Novikov, \emph{{The Hypothesis of Cores Retarded during Expansion and the Hot Cosmological Model}}, {\emph{Sov. Astron.} {\bfseries 10} (1967) 602}.

\bibitem{Carr:1974nx}
B.J.~Carr and S.W.~Hawking, \emph{{Black holes in the early Universe}}, \href{https://doi.org/10.1093/mnras/168.2.399}{\emph{Mon. Not. Roy. Astron. Soc.} {\bfseries 168} (1974) 399}.

\bibitem{Chapline:1975ojl}
G.F.~Chapline, \emph{{Cosmological effects of primordial black holes}}, \href{https://doi.org/10.1038/253251a0}{\emph{Nature} {\bfseries 253} (1975) 251}.

\bibitem{Abbott:2016blz}
{\scshape LIGO Scientific, Virgo} collaboration, \emph{{Observation of Gravitational Waves from a Binary Black Hole Merger}}, \href{https://doi.org/10.1103/PhysRevLett.116.061102}{\emph{Phys. Rev. Lett.} {\bfseries 116} (2016) 061102} [\href{https://arxiv.org/abs/1602.03837}{{\ttfamily 1602.03837}}].

\bibitem{Abbott:2020gyp}
{\scshape LIGO Scientific, Virgo} collaboration, \emph{{Population Properties of Compact Objects from the Second LIGO-Virgo Gravitational-Wave Transient Catalog}}, \href{https://doi.org/10.3847/2041-8213/abe949}{\emph{Astrophys. J. Lett.} {\bfseries 913} (2021) L7} [\href{https://arxiv.org/abs/2010.14533}{{\ttfamily 2010.14533}}].

\bibitem{KAGRA:2021vkt}
{\scshape KAGRA, VIRGO, LIGO Scientific} collaboration, \emph{{GWTC-3: Compact Binary Coalescences Observed by LIGO and Virgo during the Second Part of the Third Observing Run}}, \href{https://doi.org/10.1103/PhysRevX.13.041039}{\emph{Phys. Rev. X} {\bfseries 13} (2023) 041039} [\href{https://arxiv.org/abs/2111.03606}{{\ttfamily 2111.03606}}].

\bibitem{Aghanim:2018eyx}
{\scshape Planck} collaboration, \emph{{Planck 2018 results. VI. Cosmological parameters}}, \href{https://doi.org/10.1051/0004-6361/201833910}{\emph{Astron. Astrophys.} {\bfseries 641} (2020) A6} [\href{https://arxiv.org/abs/1807.06209}{{\ttfamily 1807.06209}}].

\bibitem{Bertone:2016nfn}
G.~Bertone and D.~Hooper, \emph{{History of dark matter}}, \href{https://doi.org/10.1103/RevModPhys.90.045002}{\emph{Rev. Mod. Phys.} {\bfseries 90} (2018) 045002} [\href{https://arxiv.org/abs/1605.04909}{{\ttfamily 1605.04909}}].

\bibitem{Cirelli:2024ssz}
M.~Cirelli, A.~Strumia and J.~Zupan, \emph{{Dark Matter}},  \href{https://arxiv.org/abs/2406.01705}{{\ttfamily 2406.01705}}.

\bibitem{Carr:2020xqk}
B.~Carr and F.~Kuhnel, \emph{{Primordial Black Holes as Dark Matter: Recent Developments}}, \href{https://doi.org/10.1146/annurev-nucl-050520-125911}{\emph{Ann. Rev. Nucl. Part. Sci.} {\bfseries 70} (2020) 355} [\href{https://arxiv.org/abs/2006.02838}{{\ttfamily 2006.02838}}].

\bibitem{Bird:2022wvk}
S.~Bird et~al., \emph{{Snowmass2021 Cosmic Frontier White Paper: Primordial black hole dark matter}}, \href{https://doi.org/10.1016/j.dark.2023.101231}{\emph{Phys. Dark Univ.} {\bfseries 41} (2023) 101231} [\href{https://arxiv.org/abs/2203.08967}{{\ttfamily 2203.08967}}].

\bibitem{Green:2024bam}
A.M.~Green, \emph{{Primordial black holes as a dark matter candidate - a brief overview}}, \href{https://doi.org/10.1016/j.nuclphysb.2024.116494}{\emph{Nucl. Phys. B} {\bfseries 1003} (2024) 116494} [\href{https://arxiv.org/abs/2402.15211}{{\ttfamily 2402.15211}}].

\bibitem{Carr:2016hva}
B.J.~Carr, K.~Kohri, Y.~Sendouda and J.~Yokoyama, \emph{{Constraints on primordial black holes from the Galactic gamma-ray background}}, \href{https://doi.org/10.1103/PhysRevD.94.044029}{\emph{Phys. Rev. D} {\bfseries 94} (2016) 044029} [\href{https://arxiv.org/abs/1604.05349}{{\ttfamily 1604.05349}}].

\bibitem{Green:2020jor}
A.M.~Green and B.J.~Kavanagh, \emph{{Primordial Black Holes as a dark matter candidate}}, \href{https://doi.org/10.1088/1361-6471/abc534}{\emph{J. Phys. G} {\bfseries 48} (2021) 043001} [\href{https://arxiv.org/abs/2007.10722}{{\ttfamily 2007.10722}}].

\bibitem{Hawking:1974rv}
S.W.~Hawking, \emph{{Black hole explosions}}, \href{https://doi.org/10.1038/248030a0}{\emph{Nature} {\bfseries 248} (1974) 30}.

\bibitem{Hawking:1974sw}
S.W.~Hawking, \emph{{Particle Creation by Black Holes}}, \href{https://doi.org/10.1007/BF02345020}{\emph{Commun. Math. Phys.} {\bfseries 43} (1975) 199}.

\bibitem{Page:1976df}
D.N.~Page, \emph{{Particle Emission Rates from a Black Hole: Massless Particles from an Uncharged, Nonrotating Hole}}, \href{https://doi.org/10.1103/PhysRevD.13.198}{\emph{Phys. Rev. D} {\bfseries 13} (1976) 198}.

\bibitem{Page:1976ki}
D.N.~Page, \emph{{Particle Emission Rates from a Black Hole. 2. Massless Particles from a Rotating Hole}}, \href{https://doi.org/10.1103/PhysRevD.14.3260}{\emph{Phys. Rev. D} {\bfseries 14} (1976) 3260}.

\bibitem{MacGibbon:2007yq}
J.H.~MacGibbon, B.J.~Carr and D.N.~Page, \emph{{Do Evaporating Black Holes Form Photospheres?}}, \href{https://doi.org/10.1103/PhysRevD.78.064043}{\emph{Phys. Rev. D} {\bfseries 78} (2008) 064043} [\href{https://arxiv.org/abs/0709.2380}{{\ttfamily 0709.2380}}].

\bibitem{Carr:2020gox}
B.~Carr, K.~Kohri, Y.~Sendouda and J.~Yokoyama, \emph{{Constraints on primordial black holes}}, \href{https://doi.org/10.1088/1361-6633/ac1e31}{\emph{Rept. Prog. Phys.} {\bfseries 84} (2021) 116902} [\href{https://arxiv.org/abs/2002.12778}{{\ttfamily 2002.12778}}].

\bibitem{Auffinger:2022khh}
J.~Auffinger, \emph{{Primordial black hole constraints with Hawking radiation\textemdash{}A review}}, \href{https://doi.org/10.1016/j.ppnp.2023.104040}{\emph{Prog. Part. Nucl. Phys.} {\bfseries 131} (2023) 104040} [\href{https://arxiv.org/abs/2206.02672}{{\ttfamily 2206.02672}}].

\bibitem{1976ApJ...206....8C}
B.J.~{Carr}, \emph{{Some cosmological consequences of primordial black-hole evaporations.}}, \href{https://doi.org/10.1086/154351}{\emph{\apj} {\bfseries 206} (1976) 8}.

\bibitem{Carr:2009jm}
B.J.~Carr, K.~Kohri, Y.~Sendouda and J.~Yokoyama, \emph{{New cosmological constraints on primordial black holes}}, \href{https://doi.org/10.1103/PhysRevD.81.104019}{\emph{Phys. Rev. D} {\bfseries 81} (2010) 104019} [\href{https://arxiv.org/abs/0912.5297}{{\ttfamily 0912.5297}}].

\bibitem{Lehoucq:2009ge}
R.~Lehoucq, M.~Casse, J.M.~Casandjian and I.~Grenier, \emph{{New constraints on the primordial black hole number density from Galactic gamma-ray astronomy}}, \href{https://doi.org/10.1051/0004-6361/200911961}{\emph{Astron. Astrophys.} {\bfseries 502} (2009) 37} [\href{https://arxiv.org/abs/0906.1648}{{\ttfamily 0906.1648}}].

\bibitem{Wright:1995bi}
E.L.~Wright, \emph{{On the density of pbh's in the galactic halo}}, \href{https://doi.org/10.1086/176910}{\emph{Astrophys. J.} {\bfseries 459} (1996) 487} [\href{https://arxiv.org/abs/astro-ph/9509074}{{\ttfamily astro-ph/9509074}}].

\bibitem{Arbey:2019mbc}
A.~Arbey and J.~Auffinger, \emph{{BlackHawk: A public code for calculating the Hawking evaporation spectra of any black hole distribution}}, \href{https://doi.org/10.1140/epjc/s10052-019-7161-1}{\emph{Eur. Phys. J. C} {\bfseries 79} (2019) 693} [\href{https://arxiv.org/abs/1905.04268}{{\ttfamily 1905.04268}}].

\bibitem{Ballesteros:2019exr}
G.~Ballesteros, J.~Coronado-Bl\'azquez and D.~Gaggero, \emph{{X-ray and gamma-ray limits on the primordial black hole abundance from Hawking radiation}}, \href{https://doi.org/10.1016/j.physletb.2020.135624}{\emph{Phys. Lett. B} {\bfseries 808} (2020) 135624} [\href{https://arxiv.org/abs/1906.10113}{{\ttfamily 1906.10113}}].

\bibitem{Laha:2020ivk}
R.~Laha, J.B.~Mu\~noz and T.R.~Slatyer, \emph{{INTEGRAL constraints on primordial black holes and particle dark matter}}, \href{https://doi.org/10.1103/PhysRevD.101.123514}{\emph{Phys. Rev. D} {\bfseries 101} (2020) 123514} [\href{https://arxiv.org/abs/2004.00627}{{\ttfamily 2004.00627}}].

\bibitem{Tan:2024nbx}
X.-h.~Tan and J.-q.~Xia, \emph{{Revisiting bounds on primordial black hole as dark matter with X-ray background}}, \href{https://doi.org/10.1088/1475-7516/2024/09/022}{\emph{JCAP} {\bfseries 09} (2024) 022} [\href{https://arxiv.org/abs/2404.17119}{{\ttfamily 2404.17119}}].

\bibitem{Boudaud:2018hqb}
M.~Boudaud and M.~Cirelli, \emph{{Voyager 1 $e^\pm$ Further Constrain Primordial Black Holes as Dark Matter}}, \href{https://doi.org/10.1103/PhysRevLett.122.041104}{\emph{Phys. Rev. Lett.} {\bfseries 122} (2019) 041104} [\href{https://arxiv.org/abs/1807.03075}{{\ttfamily 1807.03075}}].

\bibitem{Dasgupta:2019cae}
B.~Dasgupta, R.~Laha and A.~Ray, \emph{{Neutrino and positron constraints on spinning primordial black hole dark matter}}, \href{https://doi.org/10.1103/PhysRevLett.125.101101}{\emph{Phys. Rev. Lett.} {\bfseries 125} (2020) 101101} [\href{https://arxiv.org/abs/1912.01014}{{\ttfamily 1912.01014}}].

\bibitem{DeRocco:2019fjq}
W.~DeRocco and P.W.~Graham, \emph{{Constraining Primordial Black Hole Abundance with the Galactic 511 keV Line}}, \href{https://doi.org/10.1103/PhysRevLett.123.251102}{\emph{Phys. Rev. Lett.} {\bfseries 123} (2019) 251102} [\href{https://arxiv.org/abs/1906.07740}{{\ttfamily 1906.07740}}].

\bibitem{Laha:2019ssq}
R.~Laha, \emph{{Primordial Black Holes as a Dark Matter Candidate Are Severely Constrained by the Galactic Center 511 keV $\gamma$ -Ray Line}}, \href{https://doi.org/10.1103/PhysRevLett.123.251101}{\emph{Phys. Rev. Lett.} {\bfseries 123} (2019) 251101} [\href{https://arxiv.org/abs/1906.09994}{{\ttfamily 1906.09994}}].

\bibitem{DelaTorreLuque:2024qms}
P.~De~la Torre~Luque, J.~Koechler and S.~Balaji, \emph{{Refining Galactic primordial black hole evaporation constraints}}, \href{https://doi.org/10.1103/PhysRevD.110.123022}{\emph{Phys. Rev. D} {\bfseries 110} (2024) 123022} [\href{https://arxiv.org/abs/2406.11949}{{\ttfamily 2406.11949}}].

\bibitem{Halzen:1995hu}
F.~Halzen, B.~Keszthelyi and E.~Zas, \emph{{Neutrinos from primordial black holes}}, \href{https://doi.org/10.1103/PhysRevD.52.3239}{\emph{Phys. Rev. D} {\bfseries 52} (1995) 3239} [\href{https://arxiv.org/abs/hep-ph/9502268}{{\ttfamily hep-ph/9502268}}].

\bibitem{Lunardini:2019zob}
C.~Lunardini and Y.F.~Perez-Gonzalez, \emph{{Dirac and Majorana neutrino signatures of primordial black holes}}, \href{https://doi.org/10.1088/1475-7516/2020/08/014}{\emph{JCAP} {\bfseries 08} (2020) 014} [\href{https://arxiv.org/abs/1910.07864}{{\ttfamily 1910.07864}}].

\bibitem{Wang:2020uvi}
S.~Wang, D.-M.~Xia, X.~Zhang, S.~Zhou and Z.~Chang, \emph{{Constraining primordial black holes as dark matter at JUNO}}, \href{https://doi.org/10.1103/PhysRevD.103.043010}{\emph{Phys. Rev. D} {\bfseries 103} (2021) 043010} [\href{https://arxiv.org/abs/2010.16053}{{\ttfamily 2010.16053}}].

\bibitem{DeRomeri:2021xgy}
V.~De~Romeri, P.~Mart\'\i{}nez-Mirav\'e and M.~T\'ortola, \emph{{Signatures of primordial black hole dark matter at DUNE and THEIA}}, \href{https://doi.org/10.1088/1475-7516/2021/10/051}{\emph{JCAP} {\bfseries 10} (2021) 051} [\href{https://arxiv.org/abs/2106.05013}{{\ttfamily 2106.05013}}].

\bibitem{Bernal:2022swt}
N.~Bernal, V.~Mu\~noz Albornoz, S.~Palomares-Ruiz and P.~Villanueva-Domingo, \emph{{Current and future neutrino limits on the abundance of primordial black holes}}, \href{https://doi.org/10.1088/1475-7516/2022/10/068}{\emph{JCAP} {\bfseries 10} (2022) 068} [\href{https://arxiv.org/abs/2203.14979}{{\ttfamily 2203.14979}}].

\bibitem{DeRomeri:2024zqs}
V.~De~Romeri, Y.F.~Perez-Gonzalez and A.~Tolino, \emph{{Primordial black hole probes of heavy neutral leptons}}, \href{https://doi.org/10.1088/1475-7516/2025/04/018}{\emph{JCAP} {\bfseries 04} (2025) 018} [\href{https://arxiv.org/abs/2405.00124}{{\ttfamily 2405.00124}}].

\bibitem{Barrau:2001ev}
A.~Barrau, G.~Boudoul, F.~Donato, D.~Maurin, P.~Salati and R.~Taillet, \emph{{Antiprotons from primordial black holes}}, \href{https://doi.org/10.1051/0004-6361:20020313}{\emph{Astron. Astrophys.} {\bfseries 388} (2002) 676} [\href{https://arxiv.org/abs/astro-ph/0112486}{{\ttfamily astro-ph/0112486}}].

\bibitem{Barrau:2002mc}
A.~Barrau, G.~Boudoul, F.~Donato, D.~Maurin, P.~Salati, I.~Stefanon et~al., \emph{{Antideuterons as a probe of primordial black holes}}, \href{https://doi.org/10.1051/0004-6361:20021588}{\emph{Astron. Astrophys.} {\bfseries 398} (2003) 403} [\href{https://arxiv.org/abs/astro-ph/0207395}{{\ttfamily astro-ph/0207395}}].

\bibitem{Maki:1995pa}
K.~Maki, T.~Mitsui and S.~Orito, \emph{{Local flux of low-energy anti-protons from evaporating primordial black holes}}, \href{https://doi.org/10.1103/PhysRevLett.76.3474}{\emph{Phys. Rev. Lett.} {\bfseries 76} (1996) 3474} [\href{https://arxiv.org/abs/astro-ph/9601025}{{\ttfamily astro-ph/9601025}}].

\bibitem{Herms:2016vop}
J.~Herms, A.~Ibarra, A.~Vittino and S.~Wild, \emph{{Antideuterons in cosmic rays: sources and discovery potential}}, \href{https://doi.org/10.1088/1475-7516/2017/02/018}{\emph{JCAP} {\bfseries 02} (2017) 018} [\href{https://arxiv.org/abs/1610.00699}{{\ttfamily 1610.00699}}].

\bibitem{Adriani:2008zq}
O.~Adriani et~al., \emph{{A new measurement of the antiproton-to-proton flux ratio up to 100 GeV in the cosmic radiation}}, \href{https://doi.org/10.1103/PhysRevLett.102.051101}{\emph{Phys. Rev. Lett.} {\bfseries 102} (2009) 051101} [\href{https://arxiv.org/abs/0810.4994}{{\ttfamily 0810.4994}}].

\bibitem{PAMELA:2010kea}
{\scshape PAMELA} collaboration, \emph{{PAMELA results on the cosmic-ray antiproton flux from 60 MeV to 180 GeV in kinetic energy}}, \href{https://doi.org/10.1103/PhysRevLett.105.121101}{\emph{Phys. Rev. Lett.} {\bfseries 105} (2010) 121101} [\href{https://arxiv.org/abs/1007.0821}{{\ttfamily 1007.0821}}].

\bibitem{AMS:2016oqu}
{\scshape AMS} collaboration, \emph{{Antiproton Flux, Antiproton-to-Proton Flux Ratio, and Properties of Elementary Particle Fluxes in Primary Cosmic Rays Measured with the Alpha Magnetic Spectrometer on the International Space Station}}, \href{https://doi.org/10.1103/PhysRevLett.117.091103}{\emph{Phys. Rev. Lett.} {\bfseries 117} (2016) 091103}.

\bibitem{AMS:2021nhj}
{\scshape AMS} collaboration, \emph{{The Alpha Magnetic Spectrometer (AMS) on the international space station: Part II \textemdash{} Results from the first seven years}}, \href{https://doi.org/10.1016/j.physrep.2020.09.003}{\emph{Phys. Rept.} {\bfseries 894} (2021) 1}.

\bibitem{BESS:2024yma}
{\scshape BESS} collaboration, \emph{{Search for Antideuterons of Cosmic Origin Using the BESS-Polar II Magnetic-Rigidity Spectrometer}}, \href{https://doi.org/10.1103/PhysRevLett.132.131001}{\emph{Phys. Rev. Lett.} {\bfseries 132} (2024) 131001}.

\bibitem{Aramaki_2016}
T.~Aramaki, C.~Hailey, S.~Boggs, P.~von Doetinchem, H.~Fuke, S.~Mognet et~al., \emph{{Antideuteron sensitivity for the {GAPS} experiment}}, \href{https://doi.org/10.1016/j.astropartphys.2015.09.001}{\emph{\ap} {\bfseries 74} (2016) 6}.

\bibitem{GAPS:2023hag}
\emph{{GAPS contributions to the 38th International Cosmic Ray Conference (Nagoya 2023)}}, 10, 2023.

\bibitem{GAPS:2022ncd}
{\scshape GAPS} collaboration, \emph{{Sensitivity of the GAPS experiment to low-energy cosmic-ray antiprotons}}, \href{https://doi.org/10.1016/j.astropartphys.2022.102791}{\emph{Astropart. Phys.} {\bfseries 145} (2023) 102791} [\href{https://arxiv.org/abs/2206.12991}{{\ttfamily 2206.12991}}].

\bibitem{Kiraly:1981ci}
P.~Kiraly, J.~Szabelski, J.~Wdowczyk and A.W.~Wolfendale, \emph{{Anti-protons in the Cosmic Radiation}}, \href{https://doi.org/10.1038/293120a0}{\emph{Nature} {\bfseries 293} (1981) 120}.

\bibitem{Turner:1981ez}
M.S.~Turner, \emph{{Could Pbhs Be the Source of the Cosmic Ray Anti-protons?}}, \href{https://doi.org/10.1038/297379a0}{\emph{Nature} {\bfseries 297} (1982) 379}.

\bibitem{MacGibbon:1991tj}
J.H.~MacGibbon, \emph{{Quark and gluon jet emission from primordial black holes. 2. The Lifetime emission}}, \href{https://doi.org/10.1103/PhysRevD.44.376}{\emph{Phys. Rev. D} {\bfseries 44} (1991) 376}.

\bibitem{MacGibbon:1990zk}
J.H.~MacGibbon and B.R.~Webber, \emph{{Quark and gluon jet emission from primordial black holes: The instantaneous spectra}}, \href{https://doi.org/10.1103/PhysRevD.41.3052}{\emph{Phys. Rev. D} {\bfseries 41} (1990) 3052}.

\bibitem{MacGibbon:1991vc}
J.H.~MacGibbon and B.J.~Carr, \emph{{Cosmic rays from primordial black holes}}, \href{https://doi.org/10.1086/169909}{\emph{Astrophys. J.} {\bfseries 371} (1991) 447}.

\bibitem{Mitsui:1996qy}
T.~Mitsui, K.~Maki and S.~Orito, \emph{{Expected enhancement of the primary anti-proton flux at the solar minimum}}, \href{https://doi.org/10.1016/S0370-2693(96)01363-9}{\emph{Phys. Lett. B} {\bfseries 389} (1996) 169} [\href{https://arxiv.org/abs/astro-ph/9608123}{{\ttfamily astro-ph/9608123}}].

\bibitem{2012PhRvL.108e1102A}
K.~{Abe}, H.~{Fuke}, S.~{Haino}, T.~{Hams}, M.~{Hasegawa}, A.~{Horikoshi} et~al., \emph{{Measurement of the Cosmic-Ray Antiproton Spectrum at Solar Minimum with a Long-Duration Balloon Flight over Antarctica}}, \href{https://doi.org/10.1103/PhysRevLett.108.051102}{\emph{\prl} {\bfseries 108} (2012) 051102} [\href{https://arxiv.org/abs/1107.6000}{{\ttfamily 1107.6000}}].

\bibitem{Yamamoto:2013yva}
A.~Yamamoto et~al., \emph{{Search for cosmic-ray antiproton origins and for cosmological antimatter with BESS}}, \href{https://doi.org/10.1016/j.asr.2011.07.012}{\emph{Adv. Space Res.} {\bfseries 51} (2013) 227}.

\bibitem{Aramaki:2014oda}
{\scshape GAPS} collaboration, \emph{{Potential for Precision Measurement of Low-Energy Antiprotons with GAPS for Dark Matter and Primordial Black Hole Physics}}, \href{https://doi.org/10.1016/j.astropartphys.2014.03.011}{\emph{Astropart. Phys.} {\bfseries 59} (2014) 12} [\href{https://arxiv.org/abs/1401.8245}{{\ttfamily 1401.8245}}].

\bibitem{Wells:1998jv}
J.D.~Wells, A.~Moiseev and J.F.~Ormes, \emph{{Illuminating dark matter and primordial black holes with interstellar anti-protons}}, \href{https://doi.org/10.1086/307325}{\emph{Astrophys. J.} {\bfseries 518} (1999) 570} [\href{https://arxiv.org/abs/hep-ph/9811325}{{\ttfamily hep-ph/9811325}}].

\bibitem{Barrau:2002ru}
A.~Barrau, D.~Blais, G.~Boudoul and D.~Polarski, \emph{{Galactic cosmic rays from pbhs and primordial spectra with a scale}}, \href{https://doi.org/10.1016/S0370-2693(02)03060-5}{\emph{Phys. Lett. B} {\bfseries 551} (2003) 218} [\href{https://arxiv.org/abs/astro-ph/0210149}{{\ttfamily astro-ph/0210149}}].

\bibitem{Aramaki:2015pii}
T.~Aramaki et~al., \emph{{Review of the theoretical and experimental status of dark matter identification with cosmic-ray antideuterons}}, \href{https://doi.org/10.1016/j.physrep.2016.01.002}{\emph{Phys. Rept.} {\bfseries 618} (2016) 1} [\href{https://arxiv.org/abs/1505.07785}{{\ttfamily 1505.07785}}].

\bibitem{Serksnyte:2022onw}
L.~\v{S}erk\v{s}nyt\.{e} et~al., \emph{{Reevaluation of the cosmic antideuteron flux from cosmic-ray interactions and from exotic sources}}, \href{https://doi.org/10.1103/PhysRevD.105.083021}{\emph{Phys. Rev. D} {\bfseries 105} (2022) 083021} [\href{https://arxiv.org/abs/2201.00925}{{\ttfamily 2201.00925}}].

\bibitem{Korwar:2024ofe}
M.~Korwar and S.~Profumo, \emph{{Late-forming black holes and the antiproton, gamma-ray, and antihelium excesses}}, \href{https://doi.org/10.1103/PhysRevD.111.023032}{\emph{Phys. Rev. D} {\bfseries 111} (2025) 023032} [\href{https://arxiv.org/abs/2403.18656}{{\ttfamily 2403.18656}}].

\bibitem{Escriva:2022duf}
A.~Escriv\`a, F.~Kuhnel and Y.~Tada, \emph{{Primordial Black Holes}},  \href{https://arxiv.org/abs/2211.05767}{{\ttfamily 2211.05767}}.

\bibitem{Carr:1975qj}
B.J.~Carr, \emph{{The Primordial black hole mass spectrum}}, \href{https://doi.org/10.1086/153853}{\emph{Astrophys. J.} {\bfseries 201} (1975) 1}.

\bibitem{Niemeyer:1997mt}
J.C.~Niemeyer and K.~Jedamzik, \emph{{Near-critical gravitational collapse and the initial mass function of primordial black holes}}, \href{https://doi.org/10.1103/PhysRevLett.80.5481}{\emph{Phys. Rev. Lett.} {\bfseries 80} (1998) 5481} [\href{https://arxiv.org/abs/astro-ph/9709072}{{\ttfamily astro-ph/9709072}}].

\bibitem{Dolgov:1992pu}
A.~Dolgov and J.~Silk, \emph{{Baryon isocurvature fluctuations at small scales and baryonic dark matter}}, \href{https://doi.org/10.1103/PhysRevD.47.4244}{\emph{Phys. Rev. D} {\bfseries 47} (1993) 4244}.

\bibitem{Green:2016xgy}
A.M.~Green, \emph{{Microlensing and dynamical constraints on primordial black hole dark matter with an extended mass function}}, \href{https://doi.org/10.1103/PhysRevD.94.063530}{\emph{Phys. Rev. D} {\bfseries 94} (2016) 063530} [\href{https://arxiv.org/abs/1609.01143}{{\ttfamily 1609.01143}}].

\bibitem{Kannike:2017bxn}
K.~Kannike, L.~Marzola, M.~Raidal and H.~Veerm\"ae, \emph{{Single Field Double Inflation and Primordial Black Holes}}, \href{https://doi.org/10.1088/1475-7516/2017/09/020}{\emph{JCAP} {\bfseries 09} (2017) 020} [\href{https://arxiv.org/abs/1705.06225}{{\ttfamily 1705.06225}}].

\bibitem{Carr:2017jsz}
B.~Carr, M.~Raidal, T.~Tenkanen, V.~Vaskonen and H.~Veerm\"ae, \emph{{Primordial black hole constraints for extended mass functions}}, \href{https://doi.org/10.1103/PhysRevD.96.023514}{\emph{Phys. Rev. D} {\bfseries 96} (2017) 023514} [\href{https://arxiv.org/abs/1705.05567}{{\ttfamily 1705.05567}}].

\bibitem{Kuhnel:2017pwq}
F.~K\"uhnel and K.~Freese, \emph{{Constraints on Primordial Black Holes with Extended Mass Functions}}, \href{https://doi.org/10.1103/PhysRevD.95.083508}{\emph{Phys. Rev. D} {\bfseries 95} (2017) 083508} [\href{https://arxiv.org/abs/1701.07223}{{\ttfamily 1701.07223}}].

\bibitem{Dvali:2018xpy}
G.~Dvali, \emph{{A Microscopic Model of Holography: Survival by the Burden of Memory}},  \href{https://arxiv.org/abs/1810.02336}{{\ttfamily 1810.02336}}.

\bibitem{Dvali:2020wft}
G.~Dvali, L.~Eisemann, M.~Michel and S.~Zell, \emph{{Black hole metamorphosis and stabilization by memory burden}}, \href{https://doi.org/10.1103/PhysRevD.102.103523}{\emph{Phys. Rev. D} {\bfseries 102} (2020) 103523} [\href{https://arxiv.org/abs/2006.00011}{{\ttfamily 2006.00011}}].

\bibitem{Dvali:2024hsb}
G.~Dvali, J.S.~Valbuena-Berm\'udez and M.~Zantedeschi, \emph{{Memory burden effect in black holes and solitons: Implications for PBH}}, \href{https://doi.org/10.1103/PhysRevD.110.056029}{\emph{Phys. Rev. D} {\bfseries 110} (2024) 056029} [\href{https://arxiv.org/abs/2405.13117}{{\ttfamily 2405.13117}}].

\bibitem{Alexandre:2024nuo}
A.~Alexandre, G.~Dvali and E.~Koutsangelas, \emph{{New mass window for primordial black holes as dark matter from the memory burden effect}}, \href{https://doi.org/10.1103/PhysRevD.110.036004}{\emph{Phys. Rev. D} {\bfseries 110} (2024) 036004} [\href{https://arxiv.org/abs/2402.14069}{{\ttfamily 2402.14069}}].

\bibitem{Thoss:2024hsr}
V.~Thoss, A.~Burkert and K.~Kohri, \emph{{Breakdown of hawking evaporation opens new mass window for primordial black holes as dark matter candidate}}, \href{https://doi.org/10.1093/mnras/stae1098}{\emph{Mon. Not. Roy. Astron. Soc.} {\bfseries 532} (2024) 451} [\href{https://arxiv.org/abs/2402.17823}{{\ttfamily 2402.17823}}].

\bibitem{Montefalcone:2025akm}
G.~Montefalcone, D.~Hooper, K.~Freese, C.~Kelso, F.~Kuhnel and P.~Sandick, \emph{{Does Memory Burden Open a New Mass Window for Primordial Black Holes as Dark Matter?}},  \href{https://arxiv.org/abs/2503.21005}{{\ttfamily 2503.21005}}.

\bibitem{Dvali:2025ktz}
G.~Dvali, M.~Zantedeschi and S.~Zell, \emph{{Transitioning to Memory Burden: Detectable Small Primordial Black Holes as Dark Matter}},  \href{https://arxiv.org/abs/2503.21740}{{\ttfamily 2503.21740}}.

\bibitem{Ukwatta:2015iba}
T.N.~Ukwatta, D.R.~Stump, J.T.~Linnemann, J.H.~MacGibbon, S.S.~Marinelli, T.~Yapici et~al., \emph{{Primordial Black Holes: Observational Characteristics of The Final Evaporation}}, \href{https://doi.org/10.1016/j.astropartphys.2016.03.007}{\emph{Astropart. Phys.} {\bfseries 80} (2016) 90} [\href{https://arxiv.org/abs/1510.04372}{{\ttfamily 1510.04372}}].

\bibitem{Baker:2021btk}
M.J.~Baker and A.~Thamm, \emph{{Probing the particle spectrum of nature with evaporating black holes}}, \href{https://doi.org/10.21468/SciPostPhys.12.5.150}{\emph{SciPost Phys.} {\bfseries 12} (2022) 150} [\href{https://arxiv.org/abs/2105.10506}{{\ttfamily 2105.10506}}].

\bibitem{Mosbech:2022lfg}
M.R.~Mosbech and Z.S.C.~Picker, \emph{{Effects of Hawking evaporation on PBH distributions}}, \href{https://doi.org/10.21468/SciPostPhys.13.4.100}{\emph{SciPost Phys.} {\bfseries 13} (2022) 100} [\href{https://arxiv.org/abs/2203.05743}{{\ttfamily 2203.05743}}].

\bibitem{Auffinger:2022ive}
J.~Auffinger, \emph{{Primordial black holes as dark matter and Hawking radiation constraints with BlackHawk}}, Ph.D. thesis, Institut de Physique des 2 Infinis de Lyon, France, IP2I, Lyon, 2022.

\bibitem{Arbey:2021mbl}
A.~Arbey and J.~Auffinger, \emph{{Physics Beyond the Standard Model with BlackHawk v2.0}}, \href{https://doi.org/10.1140/epjc/s10052-021-09702-8}{\emph{Eur. Phys. J. C} {\bfseries 81} (2021) 910} [\href{https://arxiv.org/abs/2108.02737}{{\ttfamily 2108.02737}}].

\bibitem{Arina:2023eic}
C.~Arina, M.~Di~Mauro, N.~Fornengo, J.~Heisig, A.~Jueid and R.R.~de~Austri, \emph{{CosmiXs: cosmic messenger spectra for indirect dark matter searches}}, \href{https://doi.org/10.1088/1475-7516/2024/03/035}{\emph{JCAP} {\bfseries 03} (2024) 035} [\href{https://arxiv.org/abs/2312.01153}{{\ttfamily 2312.01153}}].

\bibitem{DiMauro:2024kml}
M.~Di~Mauro, N.~Fornengo, A.~Jueid, R.R.~de~Austri and F.~Bellini, \emph{{Nailing down the theoretical uncertainties of $\overline{\rm D}$ spectrum produced from dark matter}},  \href{https://arxiv.org/abs/2411.04815}{{\ttfamily 2411.04815}}.

\bibitem{Bierlich:2022pfr}
C.~Bierlich et~al., \emph{{A comprehensive guide to the physics and usage of PYTHIA 8.3}}, \href{https://doi.org/10.21468/SciPostPhysCodeb.8}{\emph{SciPost Phys. Codeb.} {\bfseries 2022} (2022) 8} [\href{https://arxiv.org/abs/2203.11601}{{\ttfamily 2203.11601}}].

\bibitem{Calore:2022stf}
F.~Calore, M.~Cirelli, L.~Derome, Y.~Genolini, D.~Maurin, P.~Salati et~al., \emph{{AMS-02 antiprotons and dark matter: Trimmed hints and robust bounds}}, \href{https://doi.org/10.21468/SciPostPhys.12.5.163}{\emph{SciPost Phys.} {\bfseries 12} (2022) 163} [\href{https://arxiv.org/abs/2202.03076}{{\ttfamily 2202.03076}}].

\bibitem{1990acr..book.....B}
V.S.~{Berezinskii}, S.V.~{Bulanov}, V.A.~{Dogiel} and V.S.~{Ptuskin}, \emph{{Astrophysics of cosmic rays}}, Elsevier Science and Technology (1990).

\bibitem{2002cra..book.....S}
R.~{Schlickeiser}, \emph{{Cosmic Ray Astrophysics}}, Springer (2002).

\bibitem{Strong:2007nh}
A.W.~Strong, I.V.~Moskalenko and V.S.~Ptuskin, \emph{{Cosmic-ray propagation and interactions in the Galaxy}}, \href{https://doi.org/10.1146/annurev.nucl.57.090506.123011}{\emph{Ann. Rev. Nucl. Part. Sci.} {\bfseries 57} (2007) 285} [\href{https://arxiv.org/abs/astro-ph/0701517}{{\ttfamily astro-ph/0701517}}].

\bibitem{2019PhRvD..99l3028G}
Y.~{G{\'e}nolini}, M.~{Boudaud}, P.I.~{Batista}, S.~{Caroff}, L.~{Derome}, J.~{Lavalle} et~al., \emph{{Cosmic-ray transport from AMS-02 boron to carbon ratio data: Benchmark models and interpretation}}, \href{https://doi.org/10.1103/PhysRevD.99.123028}{\emph{\prd} {\bfseries 99} (2019) 123028} [\href{https://arxiv.org/abs/1904.08917}{{\ttfamily 1904.08917}}].

\bibitem{Boudaud:2019efq}
M.~Boudaud, Y.~G\'enolini, L.~Derome, J.~Lavalle, D.~Maurin, P.~Salati et~al., \emph{{AMS-02 antiprotons' consistency with a secondary astrophysical origin}}, \href{https://doi.org/10.1103/PhysRevResearch.2.023022}{\emph{Phys. Rev. Res.} {\bfseries 2} (2020) 023022} [\href{https://arxiv.org/abs/1906.07119}{{\ttfamily 1906.07119}}].

\bibitem{Maurin:2025gsz}
D.~{Maurin} et~al., \emph{{Precision cross-sections for advancing cosmic-ray physics and other applications: a comprehensive programme for the next decade}}, {\emph{subm. to Physics Reports} (2025) } [\href{https://arxiv.org/abs/2503.16173}{{\ttfamily 2503.16173}}].

\bibitem{2001ApJ...563..172D}
F.~{Donato}, D.~{Maurin}, P.~{Salati}, A.~{Barrau}, G.~{Boudoul} and R.~{Taillet}, \emph{{Antiprotons from Spallations of Cosmic Rays on Interstellar Matter}}, \href{https://doi.org/10.1086/323684}{\emph{\apj} {\bfseries 563} (2001) 172} [\href{https://arxiv.org/abs/astro-ph/0103150}{{\ttfamily astro-ph/0103150}}].

\bibitem{2020CoPhC.24706942M}
D.~{Maurin}, \emph{{USINE: Semi-analytical models for Galactic cosmic-ray propagation}}, \href{https://doi.org/10.1016/j.cpc.2019.106942}{\emph{Computer Physics Communications} {\bfseries 247} (2020) 106942} [\href{https://arxiv.org/abs/1807.02968}{{\ttfamily 1807.02968}}].

\bibitem{Navarro:1995iw}
J.F.~Navarro, C.S.~Frenk and S.D.M.~White, \emph{{The Structure of cold dark matter halos}}, \href{https://doi.org/10.1086/177173}{\emph{Astrophys. J.} {\bfseries 462} (1996) 563} [\href{https://arxiv.org/abs/astro-ph/9508025}{{\ttfamily astro-ph/9508025}}].

\bibitem{McMillan_2016}
P.J.~McMillan, \emph{The mass distribution and gravitational potential of the milky way}, \href{https://doi.org/10.1093/mnras/stw2759}{\emph{Monthly Notices of the Royal Astronomical Society} {\bfseries 465} (2016) 76–94}.

\bibitem{2021RPPh...84j4901D}
P.F.~{de Salas} and A.~{Widmark}, \emph{{Dark matter local density determination: recent observations and future prospects}}, \href{https://doi.org/10.1088/1361-6633/ac24e7}{\emph{Reports on Progress in Physics} {\bfseries 84} (2021) 104901} [\href{https://arxiv.org/abs/2012.11477}{{\ttfamily 2012.11477}}].

\bibitem{2019AandA...625L..10G}
{Gravity Collaboration}, R.~{Abuter}, A.~{Amorim}, M.~{Baub{\"o}ck}, J.P.~{Berger}, H.~{Bonnet} et~al., \emph{{A geometric distance measurement to the Galactic center black hole with 0.3\% uncertainty}}, \href{https://doi.org/10.1051/0004-6361/201935656}{\emph{Astron. Astrophys.} {\bfseries 625} (2019) L10} [\href{https://arxiv.org/abs/1904.05721}{{\ttfamily 1904.05721}}].

\bibitem{GleesonEtAl1968a}
L.J.~{Gleeson} and W.I.~{Axford}, \emph{{Solar Modulation of Galactic Cosmic Rays}}, \href{https://doi.org/10.1086/149822}{\emph{\apj} {\bfseries 154} (1968) 1011}.

\bibitem{Fisk1971}
L.A.~{Fisk}, \emph{{Solar modulation of galactic cosmic rays, 2}}, \href{https://doi.org/10.1029/JA076i001p00221}{\emph{J. Geophys. Res.} {\bfseries 76} (1971) 221}.

\bibitem{Weinrich:2020ftb}
N.~Weinrich, M.~Boudaud, L.~Derome, Y.~Genolini, J.~Lavalle, D.~Maurin et~al., \emph{{Galactic halo size in the light of recent AMS-02 data}}, \href{https://doi.org/10.1051/0004-6361/202038064}{\emph{Astron. Astrophys.} {\bfseries 639} (2020) A74} [\href{https://arxiv.org/abs/2004.00441}{{\ttfamily 2004.00441}}].

\bibitem{Donato:2003xg}
F.~Donato, N.~Fornengo, D.~Maurin and P.~Salati, \emph{{Antiprotons in cosmic rays from neutralino annihilation}}, \href{https://doi.org/10.1103/PhysRevD.69.063501}{\emph{Phys. Rev. D} {\bfseries 69} (2004) 063501} [\href{https://arxiv.org/abs/astro-ph/0306207}{{\ttfamily astro-ph/0306207}}].

\bibitem{Genolini:2021doh}
Y.~G\'enolini, M.~Boudaud, M.~Cirelli, L.~Derome, J.~Lavalle, D.~Maurin et~al., \emph{{New minimal, median, and maximal propagation models for dark matter searches with Galactic cosmic rays}}, \href{https://doi.org/10.1103/PhysRevD.104.083005}{\emph{Phys. Rev. D} {\bfseries 104} (2021) 083005} [\href{https://arxiv.org/abs/2103.04108}{{\ttfamily 2103.04108}}].

\bibitem{Weinrich:2020cmw}
N.~Weinrich, Y.~G\'enolini, M.~Boudaud, L.~Derome and D.~Maurin, \emph{{Combined analysis of AMS-02 (Li,Be,B)/C, N/O, 3He, and 4He data}}, \href{https://doi.org/10.1051/0004-6361/202037875}{\emph{Astron. Astrophys.} {\bfseries 639} (2020) A131} [\href{https://arxiv.org/abs/2002.11406}{{\ttfamily 2002.11406}}].

\bibitem{2019PhRvL.123y1104K}
M.~{Kuhlen} and P.~{Mertsch}, \emph{{Time-Dependent AMS-02 Electron-Positron Fluxes in an Extended Force-Field Model}}, \href{https://doi.org/10.1103/PhysRevLett.123.251104}{\emph{\prl} {\bfseries 123} (2019) 251104} [\href{https://arxiv.org/abs/1909.01154}{{\ttfamily 1909.01154}}].

\bibitem{2025ApJ...982..103Z}
C.-R.~{Zhu} and K.-K.~{Duan}, \emph{{Forecasting of the Time-dependent Fluxes of Antiprotons in the AMS-02 Era}}, \href{https://doi.org/10.3847/1538-4357/adbaf1}{\emph{\apj} {\bfseries 982} (2025) 103} [\href{https://arxiv.org/abs/2501.12625}{{\ttfamily 2501.12625}}].

\bibitem{Potgieter2013}
M.~{Potgieter}, \emph{{Solar Modulation of Cosmic Rays}}, \href{https://doi.org/10.12942/lrsp-2013-3}{\emph{Living Reviews in Solar Physics} {\bfseries 10} (2013) 3} [\href{https://arxiv.org/abs/1306.4421}{{\ttfamily 1306.4421}}].

\bibitem{2025PhRvL.134e1002A}
M.~{Aguilar}, G.~{Ambrosi}, H.~{Anderson}, L.~{Arruda}, N.~{Attig}, C.~{Bagwell} et~al., \emph{{Antiprotons and Elementary Particles over a Solar Cycle: Results from the Alpha Magnetic Spectrometer}}, \href{https://doi.org/10.1103/PhysRevLett.134.051002}{\emph{\prl} {\bfseries 134} (2025) 051002}.

\bibitem{2023ApJ...947...72A}
O.P.M.~{Aslam}, X.~{Luo}, M.S.~{Potgieter}, M.D.~{Ngobeni} and X.~{Song}, \emph{{Unfolding Drift Effects for Cosmic Rays over the Period of the Sun's Magnetic Field Reversal}}, \href{https://doi.org/10.3847/1538-4357/acc24a}{\emph{\apj} {\bfseries 947} (2023) 72} [\href{https://arxiv.org/abs/2212.13397}{{\ttfamily 2212.13397}}].

\bibitem{2025arXiv250314025T}
N.~{Tomassetti}, B.~{Bertucci}, E.~{Fiandrini} and B.~{Khiali}, \emph{{Propagation Times and Energy Losses of Cosmic Protons and Antiprotons in Interplanetary Space}}, \href{https://doi.org/10.48550/arXiv.2503.14025}{\emph{arXiv e-prints} (2025) arXiv:2503.14025} [\href{https://arxiv.org/abs/2503.14025}{{\ttfamily 2503.14025}}].

\bibitem{Bess_polar_ii_antiprotons_data}
K.~Abe, H.~Fuke, S.~Haino, T.~Hams, M.~Hasegawa, A.~Horikoshi et~al., \emph{Measurement of the cosmic-ray antiproton spectrum at solar minimum with a long-duration balloon flight over antarctica}, \href{https://doi.org/10.1103/PhysRevLett.108.051102}{\emph{Phys. Rev. Lett.} {\bfseries 108} (2012) 051102}.

\bibitem{Donato:2017ywo}
F.~Donato, M.~Korsmeier and M.~Di~Mauro, \emph{{Prescriptions on antiproton cross section data for precise theoretical antiproton flux predictions}}, \href{https://doi.org/10.1103/PhysRevD.96.043007}{\emph{Phys. Rev. D} {\bfseries 96} (2017) 043007} [\href{https://arxiv.org/abs/1704.03663}{{\ttfamily 1704.03663}}].

\bibitem{Korsmeier:2018gcy}
M.~Korsmeier, F.~Donato and M.~Di~Mauro, \emph{{Production cross sections of cosmic antiprotons in the light of new data from the NA61 and LHCb experiments}}, \href{https://doi.org/10.1103/PhysRevD.97.103019}{\emph{Phys. Rev. D} {\bfseries 97} (2018) 103019} [\href{https://arxiv.org/abs/1802.03030}{{\ttfamily 1802.03030}}].

\bibitem{Arbey:2019vqx}
A.~Arbey, J.~Auffinger and J.~Silk, \emph{{Constraining primordial black hole masses with the isotropic gamma ray background}}, \href{https://doi.org/10.1103/PhysRevD.101.023010}{\emph{Phys. Rev. D} {\bfseries 101} (2020) 023010} [\href{https://arxiv.org/abs/1906.04750}{{\ttfamily 1906.04750}}].

\bibitem{Boluna:2023jlo}
X.~Boluna, S.~Profumo, J.~Bl\'e and D.~Hennings, \emph{{Searching for Exploding black holes}}, \href{https://doi.org/10.1088/1475-7516/2024/04/024}{\emph{JCAP} {\bfseries 04} (2024) 024} [\href{https://arxiv.org/abs/2307.06467}{{\ttfamily 2307.06467}}].

\bibitem{HESS:2023zzd}
{\scshape H.E.S.S.} collaboration, \emph{{Search for the evaporation of primordial black holes with H.E.S.S.}}, \href{https://doi.org/10.1088/1475-7516/2023/04/040}{\emph{JCAP} {\bfseries 04} (2023) 040} [\href{https://arxiv.org/abs/2303.12855}{{\ttfamily 2303.12855}}].

\bibitem{Donato:1999gy}
F.~Donato, N.~Fornengo and P.~Salati, \emph{{Anti-deuterons as a signature of supersymmetric dark matter}}, \href{https://doi.org/10.1103/PhysRevD.62.043003}{\emph{Phys. Rev. D} {\bfseries 62} (2000) 043003} [\href{https://arxiv.org/abs/hep-ph/9904481}{{\ttfamily hep-ph/9904481}}].

\bibitem{Donato:2008yx}
F.~Donato, N.~Fornengo and D.~Maurin, \emph{{Antideuteron fluxes from dark matter annihilation in diffusion models}}, \href{https://doi.org/10.1103/PhysRevD.78.043506}{\emph{Phys. Rev. D} {\bfseries 78} (2008) 043506} [\href{https://arxiv.org/abs/0803.2640}{{\ttfamily 0803.2640}}].

\bibitem{Korsmeier:2017xzj}
M.~Korsmeier, F.~Donato and N.~Fornengo, \emph{{Prospects to verify a possible dark matter hint in cosmic antiprotons with antideuterons and antihelium}}, \href{https://doi.org/10.1103/PhysRevD.97.103011}{\emph{Phys. Rev. D} {\bfseries 97} (2018) 103011} [\href{https://arxiv.org/abs/1711.08465}{{\ttfamily 1711.08465}}].

\bibitem{Cirelli:2014qia}
M.~Cirelli, N.~Fornengo, M.~Taoso and A.~Vittino, \emph{{Anti-helium from Dark Matter annihilations}}, \href{https://doi.org/10.1007/JHEP08(2014)009}{\emph{JHEP} {\bfseries 08} (2014) 009} [\href{https://arxiv.org/abs/1401.4017}{{\ttfamily 1401.4017}}].

\bibitem{Carlson:2014ssa}
E.~Carlson, A.~Coogan, T.~Linden, S.~Profumo, A.~Ibarra and S.~Wild, \emph{{Antihelium from Dark Matter}}, \href{https://doi.org/10.1103/PhysRevD.89.076005}{\emph{Phys. Rev. D} {\bfseries 89} (2014) 076005} [\href{https://arxiv.org/abs/1401.2461}{{\ttfamily 1401.2461}}].

\bibitem{ARAMAKI20161}
T.~Aramaki, S.~Boggs, S.~Bufalino, L.~Dal, P.~{von Doetinchem}, F.~Donato et~al., \emph{Review of the theoretical and experimental status of dark matter identification with cosmic-ray antideuterons}, \href{https://doi.org/https://doi.org/10.1016/j.physrep.2016.01.002}{\emph{Physics Reports} {\bfseries 618} (2016) 1}.

\end{thebibliography}\endgroup

\end{document}